\documentclass[useAMS,usenatbib]{mn2e}
\usepackage{psfig,graphicx,listings}  
\usepackage{float}
\usepackage{amstext}
\usepackage{amssymb}
\usepackage{amsmath, float}
\usepackage{graphicx}
\usepackage{txfonts}
\usepackage{amsmath,mathtools,revsymb,amssymb}
\usepackage{booktabs,multirow,calc,xspace,rotating}

\newcommand       \be           {\begin{equation}}
\newcommand       \ee           {\end{equation}}
\newcommand       \bea          {\begin{eqnarray}}
\newcommand       \eea          {\end{eqnarray}}
\newcommand       \Me		{M_{\rm env}}

\newcommand       \ve		{v_{\rm esc}}

\newcommand       \Lw		{L_{\rm wind}}
\newcommand       \Ledd		{L_{\rm Edd}}
\newcommand	     \rh		{r_{\rm h}}
\newcommand       \gammae		{\gamma_{\rm env}}
\newcommand       \Ke		{K_{\rm env}}
\newcommand       \Kw		{K_{\rm wind}}
\newcommand       \vc		{v_{\rm crit}}

\newcommand       \kms		{\,{\rm km \,\, s}^{-1}}

\newcommand       \mspy 	{\,{\rm M_\odot \, yr^{-1}}}

\newcommand       \K		{\,{\rm K }}

\newcommand       \erg		{\,{\rm erg }}

\newcommand\plotone[1]
 {\centering \leavevmode \includegraphics[width={0.99\columnwidth}]{#1}}

\begin{document}

\title[Super-Eddington Stellar Winds]{Super-Eddington Stellar Winds Driven by Near-Surface Energy Deposition}
\author[E. Quataert, R. Fern\'andez, D. Kasen, H. Klion \& B. Paxton]{Eliot Quataert$^{1}$, Rodrigo Fern\'andez$^{1}$, Daniel Kasen$^{1,2}$, Hannah Klion$^{1}$, \& Bill Paxton$^{3}$ \\
  $^{1}$Astronomy and Physics Departments and Theoretical Astrophysics
 Center, University of California, Berkeley,  Berkeley CA, 94720\\ 
    $^{2}$Nuclear Science Division, Lawrence Berkeley National Laboratory, Berkeley, CA 94720\\ 
       $^{3}$Kavli Institute for Theoretical Physics, University of California, Santa Barbara, CA 93106 \\ }

\maketitle
\begin{abstract}
We develop analytic and numerical models of the properties of
super-Eddington stellar winds, motivated by phases in stellar
evolution when super-Eddington energy deposition (via, e.g., unstable
fusion, wave heating, or a binary companion) heats a region near the
stellar surface.  This appears to occur in luminous blue variables (LBVs),
Type IIn supernovae progenitors, classical novae, and X-ray bursts.
We show that when the wind kinetic power exceeds Eddington, the
photons are trapped and behave like a fluid.  Convection does not play
a significant role in the wind energy transport.  The wind properties
depend on the ratio of a characteristic speed in the problem $\vc \sim
(\dot E G)^{1/5}$ (where $\dot E$ is the heating rate) to the stellar
escape speed near the heating region $\ve(r_h)$.  For $\vc \gtrsim
\ve(r_h)$ the wind kinetic power at large radii $\dot E_w \sim \dot
E$.  For $\vc \lesssim \ve(r_h)$, most of the energy is used to unbind
the wind material and thus $\dot E_w \lesssim \dot E$.
Multidimensional hydrodynamic simulations without radiation diffusion
using FLASH and one-dimensional hydrodynamic simulations with
radiation diffusion using MESA are in good agreement with the analytic
predictions.  The photon luminosity from the wind is itself
super-Eddington but in many cases the photon luminosity is likely
dominated by `internal shocks' in the wind.  We discuss the
application of our models to eruptive mass loss from massive stars and
argue that the wind models described here can account for the broad
properties of LBV outflows and the enhanced mass loss in the
years  prior to Type IIn core-collapse supernovae.
\end{abstract}

\vspace{-0.7cm}
\section{Introduction}
\label{sec:int}
\vspace{-0.1cm}

During most phases of stellar evolution, the structure of stars can stably adjust so that energy generation in the stellar interior is less than or of order the Eddington luminosity, ensuring that the star is on average in thermal and hydrostatic equilibrium.   However, this balance can be upset by instabilities or external perturbations (e.g., a binary) leading to epochs of super-Eddington energy generation.  The canonical example of this phenomena is runaway thermonuclear fusion due to either fusion under degenerate conditions (e.g., \citealt{Mestel1952}) or gas pressure dominated thin shell fusion (e.g., \citealt{Schwarzschild1965}).   This occurs e.g., during the He shell flash in low mass stars, He shell fusion on the Asymptotic Giant Branch,  classical novae,  radius expansion X-ray bursts on accreting neutron stars, and Type Ia supernovae.   

In contrast to the case of runaway fusion, however, the properties of super-Eddington heating at a given location in a star need not be solely determined by the stellar conditions at that location.  Instead, energy can effectively be deposited at a given radius by an external source (e.g., via tidal heating or via a companion during common envelope evolution; e.g., \citealt{Paczynski1976}) or by non-local redistribution of energy (e.g., via wave transport of energy in stellar interiors; e.g., \citealt{Piro2011,quataert2012}).   For this reason, we shall largely use the terms `energy generation' and `energy deposition' synonymously in this paper.

There is strong evidence that massive stars undergo periods of super-Eddington energy generation/deposition, although the physical causes are much less well understood.   Luminous blue variables (LBVs) such as Eta Carinae radiate a photon luminosity significantly exceeding the Eddington luminosity for months-decades (many dynamical times) and drive a wind whose time-averaged kinetic power exceeds both the Eddington luminosity and probably the photon luminosity (\citealt{Smith2003}; see, e.g., \citealt{Davidson2012} for a review of LBVs).    Such outbursts may dominate the total mass loss from massive stars (e.g., \citealt{Smith2006,Kochanek2011}).    Moreover,  $\sim10\%$ of  supernova (SN) progenitors  experience enhanced mass loss in the decades to weeks prior to core collapse  (much larger than can be explained by line driven winds).  Evidence for this powerful mass loss includes observations of luminous outbursts that precede supernovae  \citep{pastorello2007, foley2007, mauerhan2013,ofek2013} and mass-loss rates $\sim 10^{-3} - 1 \mspy$ inferred from observations of circumstellar interaction in Type IIn SNe (e.g., \citealt{kiewe2012,Smith2014}).

What is the response of a star to super-Eddington energy generation/deposition?   There are essentially three regimes.    If sustained energy deposition leads to heating on a timescale short compared to the local dynamical time, a propagating shock will form \citep{Dessart2010}.   By contrast, if the energy deposition occurs on a timescale long compared to the dynamical time, the response is at least initially largely hydrostatic:  the increase in thermal energy slowly lifts matter out to larger radii and convection sets in in response to photons being unable to carry the energy (e.g., \citealt{Joss1973}).   Finally, if the super-Eddington energy generation occurs for sufficiently long and/or sufficiently close to the stellar photosphere,  it can drive a powerful wind (e.g., \citealt{Kato1994,Owocki2004}).   In this paper, we are interested in the latter regime.   This requires that the energy  deposition  is on for several thermal times so that the entire star exterior to the heating region inevitably adjusts its structure significantly in response to the energy deposition.  It is important to stress that this is not always  the case.   For example, during the Helium flash in low mass stars, runaway fusion produces a highly super-Eddington energy generation rate that locally drives vigorous convection.  However, the expansion of the exterior of the star in response to the energy input from fusion lifts the degeneracy, quenching the fusion.  There is no noticeable increase in the surface luminosity (it in fact decreases slightly; e.g., \citealt{Bildsten2012}), let alone a powerful wind.   

Previous work on super-Eddington stellar winds has in many cases approached the problem as a radiative transfer problem.   For example, increases in opacity at specific temperatures (e.g., the iron opacity bump) can produce a locally super-Eddington flux and potentially drive a wind \citep{Kato1994,Eichler1995,Heger1996}.   Alternatively, \citet{shaviv2001,Owocki2004,vanMarle2008} argued that super-Eddington winds are regulated by how radiation diffuses  through a highly porous stellar atmosphere that is envisioned to be the outcome of some combination of convection, (magneto)-hydrodynamic instabilities, and/or the wind launching process itself.    In this paper, we show that when the wind kinetic power exceeds the Eddington luminosity, the photon diffusion time is long compared to the advection time in the wind (\S \ref{sec:adiabatic}):  photons are trapped and behave like a fluid.  As a result,  we argue that how the radiation diffuses through the wind/stellar atmosphere  is  less dynamically important than suggested in previous calculations.   Instead, super-Eddington winds are essentially hydrodynamic, more analogous to \citet{Parker1958}'s thermal solar wind model (generalized to a radiation pressure dominated fluid) than standard radiation-driven winds from massive stars. 

\vspace{-0.4cm}
\subsection{Outline of This Paper}

We study the response of a star to continuous energy deposition near the surface at a rate larger than the typical Eddington luminosity.   Throughout much of the paper, we are relatively agnostic as to the physical origin of this heating.  It could represent, e.g., the outcome of unstable nuclear fusion or heating via waves generated in the core that carry a large energy flux to the surface where the waves dissipate or heating triggered by a binary companion (e.g., tides, common envelope).

As noted above, the response of a star to super-Eddington energy deposition has been studied extensively in the context of runaway fusion in stars.   This work has tended to focus on the limit in which sustained heating for of order a thermal time or more leads to the formation of an extended convection zone that carries the excess energy to larger radii (e.g., \citealt{Woosley2004,Weinberg2006,Piro2008}).   We briefly review this quasi-hydrostatic response in \S \ref{sec:conv}.    We then discuss the role of convective energy transport in wind solutions, rather than hydrostatic envelopes (\S \ref{sec:conv-analytics}).   We show that energy transport by convection is subdominant in steady state winds.   Thus although convection can  play an important role inflating a stellar envelope in response to energy deposition, it becomes unimportant once a roughly steady wind develops.  This typically occurs on a few thermal times.    

In \S \ref{sec:wind-analytics} we introduce and solve a simple model problem for  super-Eddington winds neglecting energy transport by convection and radiation.   These are radiation driven winds, in a regime where the photons are trapped and behave like a fluid.     \S \ref{sec:numerics}  presents  numerical examples of wind generation using both FLASH and  MESA  that are in good agreement with the analytic models.   \S \ref{sec:observables} presents some of the observational properties of super-Eddington stellar winds.   Finally, \S \ref{sec:applications} discusses several  applications of our work to massive stars and \S \ref{sec:disc} summarizes our main conclusions and highlights some key questions not addressed by our models.   

A central feature of all of the models in this paper is that we assume that an unspecified process (e.g., fusion, wave dissipation) leads to heating at a rate $\dot E$ near a heating radius radius $\rh$.    This  heating is put in `by hand' in our calculations.     The advantage of this treatment is that it means that our model is potentially  applicable to a range of physical processes and stellar contexts.   However, when scaling our analytic and numerical results, we focus on the application to super-Eddington winds from massive stars.    It is also important to stress that the photon luminosity is not an input quantity in our model, as is often the case when super-Eddington wind models are formulated in terms of what wind properties result from a given super-Eddington luminosity (or, equivalently, a given Eddington ratio $\Gamma$).  Instead, the photon luminosity is a derived quantity in our model, which depends on the heating rate $\dot E$ and the properties of the resulting wind (\S \ref{sec:observables}).  We believe that this is a more appropriate formulation of the problem of super-Eddington stellar winds for the systems of interest in this paper.   

\vspace{-0.7cm}
\section{Hydrostatic Adjustment in Response to Energy Deposition}
\label{sec:conv}
\vspace{-0.1cm}

Assume that  there is some source of heating at a rate $\dot E$ near a heating  radius $\rh$.The local thermal energy  increases in response to the heating  on the timescale $\sim 4 \pi \rh^2 u H/\dot E$ where $u$ and $H$ are the thermal energy per unit volume and scale-height at $\rh$, respectively.    We assume throughout that the source of heating is on for multiple thermal times so that the outer envelope of the star inevitably adjusts its structure significantly in response to the energy deposition. 

After of order one local thermal time, the stellar envelope begins to expand outwards due to the increased thermal pressure.  Neglecting for now the possibility of a wind, super-Eddington energy deposition also inevitably drives convection, which attempts to carry the energy that photons cannot (e.g., \citealt{Joss1973}).   In hydrostatic models (i.e., absent winds), this leads to the creation of a large convective envelope even if the star was initially compact.   The total timescale to rearrange the structure of the stellar envelope is of order the thermal time of all of the mass exterior to $\sim \rh$:
\be
t_{\rm thermal} \simeq \frac{G M(<\rh) M(>r_h)}{2 \, \rh \, \dot E}
\label{eq:thermal}
\ee
where $M(<\rh)$ is the stellar mass within $\rh$ and $M(>r_h)$ is the stellar mass exterior to $\rh$.    The thermal time as defined in equation \ref{eq:thermal} is often significantly longer than the local thermal time at $\rh$ because all of the stellar envelope participates in the convective and hydrostatic readjustment.

\begin{figure}
\resizebox{9cm}{!}{\plotone{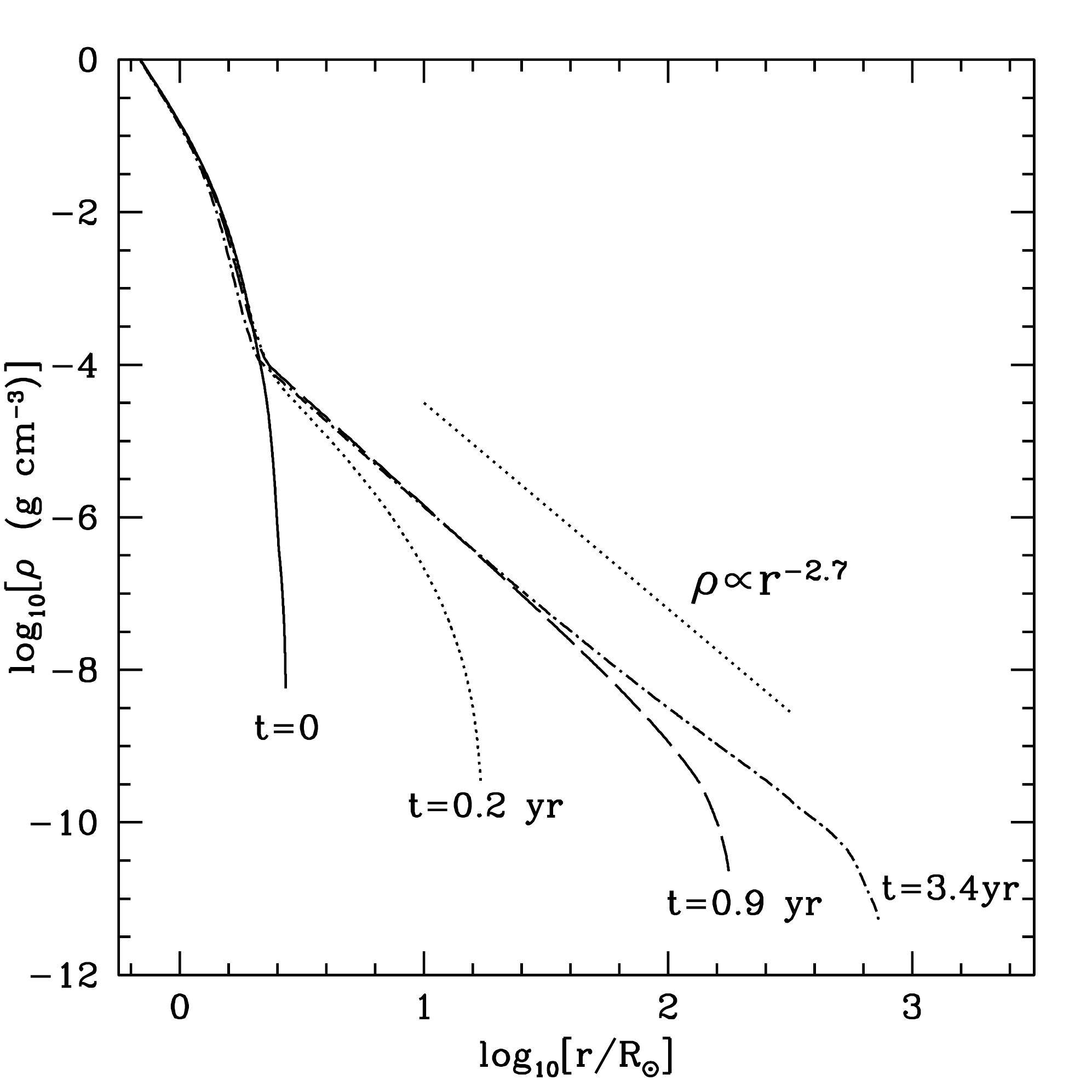}}
\caption{Density profile as a function of radius at different times after the onset of energy deposition with $\dot E = 3 \times 10^6 L_\odot$ at $r_h \simeq 2 R_\odot$ (in an 11.1 $M_\odot$, $Z = Z_\odot$ model at core He exhaustion).  These are {hydrostatic} MESA models that cannot develop a wind.   The initial thermal time (eq. \ref{eq:thermal}) at $r_h$ is $\simeq 0.2$ yr, which sets the initial expansion time of the envelope.  At late times the heating has generated a spatially extended convective envelope with a density profile similar to the $\rho \propto r^{-3}$ profile expected for radiation dominated convection.   Compare with Figure \ref{fig:rho-hydro} which shows analogous density profiles in hydrodynamic MESA models that do develop a wind.}
\label{fig:density}
\end{figure}

\begin{figure}
\resizebox{9cm}{!}{\plotone{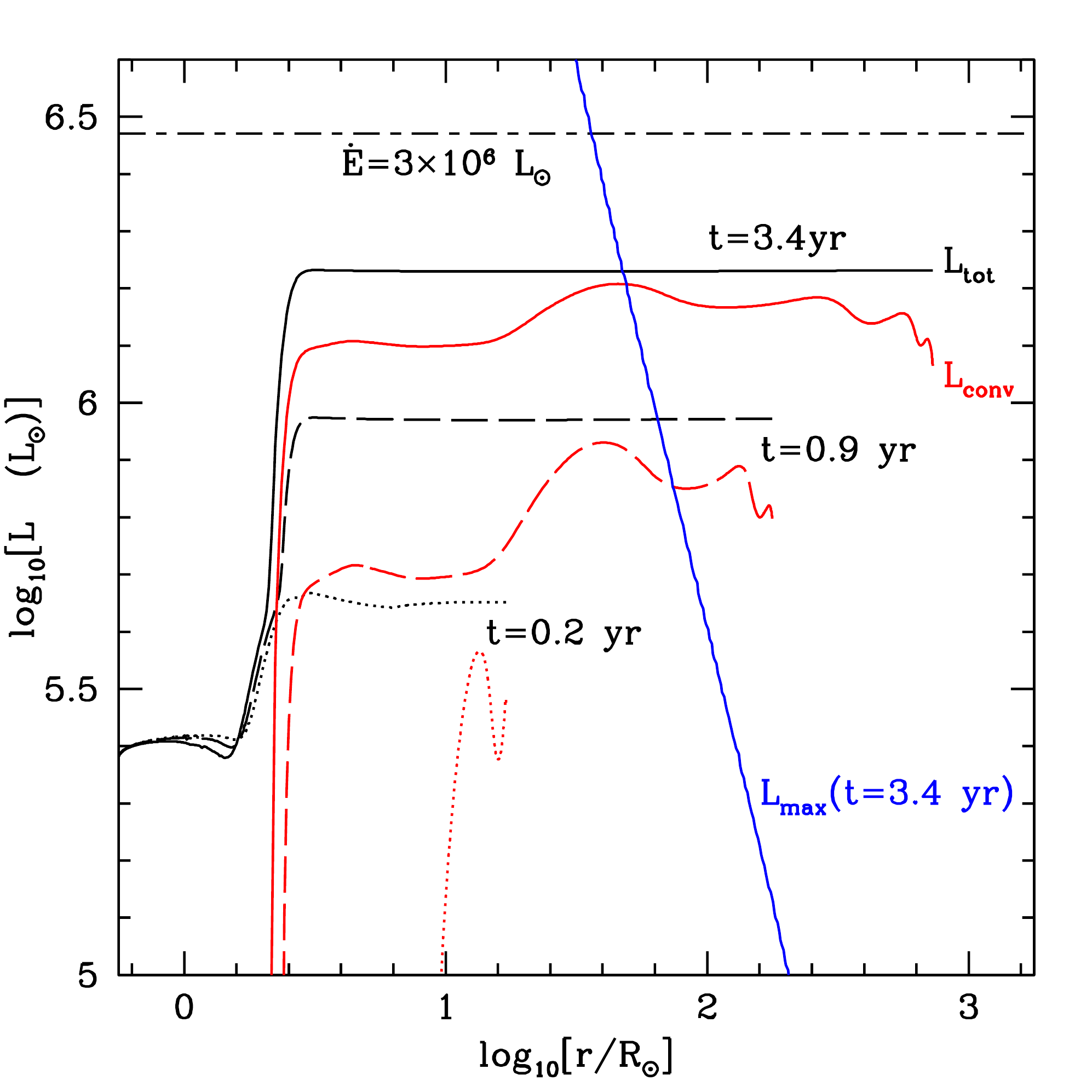}}
\caption{Convective ($L_{\rm conv}$) and total ($L_{\rm tot}$, sum of radiative and convective) luminosities as a function of radius at different times after the onset of energy deposition with $\dot E = 3 \times 10^6 L_\odot$ at $r_h \simeq 2 R_\odot$ (in an 11.1 $M_\odot$, $Z = Z_\odot$ model at core He exhaustion).  These are {hydrostatic} MESA models that cannot develop a wind.   The initial thermal time (eq. \ref{eq:thermal}) at $r_h$ is $\simeq 0.2$ yr, which sets the initial expansion time of the envelope.  At late times most of the energy is carried to large radii by convection.   $L_{max} = 4 \pi r^2 \rho c_s^3$ is the maximal convective power for subsonic convection (not applicable in these models because of the use MLT++ in MESA; see \S \ref{sec:conv}). Compare with Figure \ref{fig:lum-hydro} which shows analogous luminosity profiles in hydrodynamic MESA models that do develop a wind.}
\label{fig:lum}
\end{figure}

To quantitatively illustrate this process, we have carried out a number of calculations using the MESA stellar evolution code \citep{mesaI,mesaII,mesaIII} in which we  inject energy into the star at a specified rate.   The inlists for the models in this section are given in Appendix \ref{sec:mesa1}.  The two key properties of these models are, first, that they are essentially hydrostatic, so that there is no option for a wind to develop.   Secondly, for simplicity we utilize the MLT++ option in MESA which forces the convection to be efficient, i.e., to have roughly constant entropy, in radiation pressure dominated regions \citep{mesaII}.    This can formally become a poor assumption at large radii, but we show in the next section that once a wind develops convection in fact plays little role in setting the wind properties.  
 
We focus here on an initially compact blue supergiant (BSG) because this highlights most dramatically how the stellar structure adjusts in response to energy deposition in the envelope.   The stellar model is derived from a $30 M_\odot$, $Z = Z_\odot$ star evolved to He core exhaustion without energy deposition.  At He core exhaustion, the stellar mass and radius are $11.1 M_\odot$ and $2.7 R_\odot$, respectively.   We then deposit energy at a rate $\dot E = 3 \times 10^6 L_\odot$ in a region centered at $\rh = 2 R_\odot$.  This choice of $\dot E$ is approximately 10 times the electron-scattering Eddington luminosity.   The  stellar mass exterior to $\rh$ is initially $M(>r_h) \sim 3 \times 10^{-3} M_\odot$ although this increases to $M(>r_h) \sim 0.03 M_\odot$ during the hydrostatic readjustment.   

Figure \ref{fig:density} shows  the density profile of the star at several different times as it expands from the initially compact BSG to become a red supergiant (RSG).   The expansion occurs on the thermal time in equation \ref{eq:thermal}: this is $\simeq 0.2$ yr for the initial BSG model but the entire process of inflating the stellar envelope takes about a factor of $\sim 10$ longer as more mass expands out and is incorporated into the convective envelope.  The density profile at large radii in the RSG configuration is well described by a power-law.   This is very close to a $\gamma = 4/3$ polytropic atmosphere expected for efficient convection in a radiation pressure dominated fluid (as we derive  analytically in the next section).

Figure \ref{fig:lum} shows the total (radiative and convective) $L_{\rm tot}$ (black) and convective $L_{\rm conv}$ (red) luminosities as a function of radius for the same models shown in Figure \ref{fig:density}.  For comparison, we also show the energy deposition rate of $\dot E= 3 \times 10^6 L_\odot$.   Recall that the energy deposition occurs at $\rh \simeq 2 R_\odot$.   This is why $L_{\rm tot}$ and $L_{\rm conv}$ rise abruptly at that radius.    Figure \ref{fig:lum} shows that at early times ($t = 0.2$ yr), very little of the energy deposited into the star has gone into photon or convective power (since $L_{\rm tot}, L_{\rm conv} \ll \dot E$).  Instead, most of the energy  has gone into heat and P-dV work to expand the envelope.   At later times, however, the star has reached a new equilibrium in which much of the energy supplied at small radii is carried to large radii via convection and, to a lesser extent, photon diffusion.

\vspace{-0.7cm}
\section{Analytic Models of Super-Eddington Winds}
\label{sec:analytics}
\vspace{-0.1cm}

In this section we derive the properties of steady state spherically symmetric radiation-pressure driven winds in response to super-Eddington energy deposition.   We first show that convection is not  an important energy transport mechanism in such winds, in spite of its role in initially expanding the stellar envelope in response to energy deposition (as shown in \S \ref{sec:conv}).   We solve a model steady state spherically symmetric wind problem in \S \ref{sec:wind-analytics} and describe how the wind properties depend on the energy deposition rate $\dot E$ and the properties of the stellar model.   Throughout we neglect diffusive transport of energy and assume that the optical depths are sufficiently high that the photons are trapped and so behave like a $\gamma = 4/3$ fluid.  Physically, this assumption is motivated by the fact that photon diffusion can only transport energy at the Eddington luminosity and yet we are interested in problems for which the energy deposition rate  is super-Eddington.   We check the validity of this assumption in \S \ref{sec:adiabatic} and show that it requires that the wind kinetic energy flux exceed the Eddington luminosity.   In \S \ref{sec:Gamma} we briefly contrast our models with super-Eddington wind models formulated in terms of the Eddington ratio $\Gamma$.

\subsection{The Role of Convection in Wind Solutions}
\label{sec:conv-analytics}

The blue line in Figure \ref{fig:lum} shows the maximum power that subsonic convection can carry for the hydrostatic MESA model at t = 3.4 yr.  The convective energy transport should physically be limited by the fact that the velocities  remain subsonic, so that
\be
L_{\rm max} \sim 4 \pi r^2 \rho c_s^3.
\label{eq:lmax}
\ee
Figure \ref{fig:lum} shows that the maximum convective power decreases rapidly at large radii and by $r \sim 50 R_\odot$ convection would have difficulty carrying the required energy flux (this is not true in the numerical models because of MLT++ in MESA).   Assuming the convective power $L_{\rm conv} \sim \dot E$ is super-Eddington, photons also cannot carry the energy outwards.   An a priori plausible hypothesis is that because of the failure of either convection or photons to carry the energy $\dot E$ outwards, the pressure will build up  at roughly  the location where convection ceases to efficiently carry the energy outwards.  The radiation pressure gradient would then accelerate the matter outwards in a wind.   We now show, however, that this hypothesized transition cannot occur in a steady state wind and instead convective energy transport is sub-dominant throughout the wind.

Consider a hypothetical wind in which the flow is heated near a radius $\rh$ at a rate per unit volume $\dot q$ (with $\int dr \, 4 \pi r^2 \dot q = \dot E$) and then driven outwards by radiation pressure at radii much greater than the heating radius $\rh$ because of a failure of convection to transport the energy outwards.   Neglecting radiation transport of energy, the energy equation describing such a flow in steady state and spherical symmetry would be
\be
\rho v T \frac{ds}{dr} = \dot q -\nabla \cdot F_{\rm conv} = \dot q -\frac{1}{4 \pi r^2} \frac{d L_{\rm conv}}{dr}.
\label{eq:energy}
\ee
At large radii $\gg \rh$, $\dot q = 0$ by assumption.   If the failure of convection to transport energy outwards is to drive a wind then $d L_{\rm conv}/dr < 0$ as energy is transferred from convection to thermal energy and then to a wind.   This implies, however, that the right-hand side of equation \ref{eq:energy} is $>0$ and thus that $ds/dr > 0$, i.e., the flow is convectively stable.   This demonstrates that there is a not a consistent steady state solution in which the flow transitions with radius from a large hydrostatic convective envelope to a wind.   
Moreover, once a steady wind is established, convection is not an important source of energy transport.   To see the latter, again consider equation \ref{eq:energy} but now with $F_{\rm conv} = 0$.     In the heating region $\dot q > 0$, $ds/dr > 0$ and so the flow is convectively stable.  Outside the heating region, $\dot q = 0$ and the flow is buoyantly neutral.  

The above considerations demonstrate that although convection does play a role in initially expanding the stellar envelope in response to energy deposition (see \S \ref{sec:conv}), convection is unimportant once a steady  wind is established.    For these reasons, we do not consider convective energy transport in our wind solutions that follow.   The numerical models in \S \ref{sec:numerics} confirm the sub-dominance of convection in the wind dynamics.

\subsection{Spherically Symmetric Super-Eddington Winds}
\label{sec:wind-analytics}

We consider a simplified model problem to describe the physics of  stellar winds with super-Eddington energy deposition.  The structure of the stellar envelope can be modeled analytically if we neglect its self-gravity.   We take $P = \Ke \rho^{\gammae}$, where $\Ke$ is an effective entropy of the envelope and $\gammae$ quantifies its stratification.  Solving hydrostatic equilibrium, $dP/dr = -G\rho M/r^2$, where $M$ is the total stellar mass interior to the stellar envelope, yields the density and sound speed profiles:
\be
\rho_{\rm env}(r) = \left(\frac{\gammae-1}{\gammae \Ke}\right)^{1/(\gammae-1)} \left(\frac{GM}{r}-\frac{GM}{R}\right)^{1/(\gammae-1)}
\label{eq:rho}
\ee
and
\be
c_s^2(r) \equiv \frac{P}{\rho} = \frac{\gammae-1}{\gammae} \left(\frac{GM}{r}-\frac{GM}{R}\right).
\label{eq:csq}
\ee
The stellar radius $R$ in equations \ref{eq:rho} \& \ref{eq:csq} is defined to be where $\rho = 0$.    Note also that in equation \ref{eq:csq} the sound speed $c_s$ is defined without a factor of the adiabatic index, so it is more analogous to the isothermal sound speed.   We use this definition throughout the paper.

The analytic density profile in equation \ref{eq:rho} for a radiation dominated convective envelope with $\gammae = 4/3$  becomes $\rho \propto r^{-3}$, which is in good agreement with the numerical results in Figure \ref{fig:density} for $t = 0.9$ and 3.4 yr, i.e., once $R \gg \rh$.   The small difference in power law slope is because in the numerical model gas pressure contributes about 10\% of the total pressure, which modifies the adiabatic index from the pure radiation pressure value.

Given equation \ref{eq:rho} as the background density profile of the stellar envelope, we solve the steady state spherically symmetric wind equations for a wind subject to a total heating rate $\dot E$ at radius $r_h$.   In the analytic calculation it is convenient to assume that the width of the heating region is $\ll r_h$ so that the heating is relatively spatially localized.   Mass, momentum, and energy conservation for a wind in thermal equilibrium (equal radiation and gas temperatures) can be written as
\be
\dot M = 4 \pi r^2 \rho v = {\rm constant}
\label{eq:mdot}
\ee
\be
v \frac{dv}{dr} = -\frac{1}{\rho} \frac{dP}{dr} - \frac{GM}{r^2}
\label{eq:mom}
\ee
\be
 \frac{d}{dr}\left(L_{\rm rad} + \dot M \, Be \right) = \frac{d}{dr}\left[L_{\rm rad} + \dot M \left(\frac{1}{2}v^2 + h - \frac{G M}{r}\right) \right] = 4 \pi r^2 \dot q,
\label{eq:energy2}
\ee
where $L_{\rm rad}$ is the total power carried by photons.  $Be$ in equation \ref{eq:energy2} is the Bernoulli parameter, the conserved energy per unit mass flux  for a steady wind absent radiation, and $h$ is the enthalpy of the fluid.

To solve equations \ref{eq:mdot}-\ref{eq:energy2} analytically, we make several simplifying approximations.  First, we neglect $L_{\rm rad}$ because we are interested in systems with super-Eddington energy  deposition.  We assess the validity of this approximation in \S \ref{sec:adiabatic}.      We further assume that outside the heating radius, the fluid is radiation dominated with an adiabatic index $\gamma = 4/3$.    This is appropriate even if the stellar envelope prior to heating has a stratification $\gammae \ne 4/3$.  The reason is that if the net heating in the heating region is super-Eddington it will increase the entropy to the point where radiation pressure  dominates.    For a fluid with adiabatic index $\gamma = 4/3$, the enthalpy in equation \ref{eq:energy2} is given by $h = 4 \, c_s^2$, where $c_s^2 \equiv P/\rho$  and $P$ is the radiation pressure. 

Equations  \ref{eq:energy} and \ref{eq:energy2} are equivalent formulations of the steady state, spherically symmetric energy equation, but the latter is more convenient  for the wind problem.   The reason is that equation \ref{eq:energy2} can be integrated to yield global conservation of energy in the wind
\be
\dot M \left[Be(r_h)-Be(r_0)\right] = \dot M \left[Be(r \rightarrow \infty)-Be(r_0)\right] = \dot E
\label{eq:energyBe}
\ee
where $Be(r_h)$ is assumed to be evaluated just outside the heating region and $r_o$ is a radius in the stellar envelope just interior to where the heating occurs.  The first two expressions in equation \ref{eq:energyBe} are equivalent because $\dot q$ is assumed to be zero outside $\sim r_h$ and hence energy is conserved between $r_h$ and large radii.    $Be(r_o)$ in equation \ref{eq:energyBe} is effectively the binding energy per unit mass of the stellar envelope prior to heating.  We define
\be
Be(r_o) \equiv - \frac{f}{2} \ve(r_h)^2.
\label{eq:Berh}
\ee
Physically, equation \ref{eq:Berh} corresponds to the assumption that a typical speed of order the escape speed at the heating radius is required to unbind matter from the stellar envelope.  The dimensionless parameter $f$ quantifies the binding energy of matter in the stellar envelope and depends on the exact structure of the  envelope.   For example, for a  polytropic atmosphere with negligible mass (so that self-gravity can be neglected) described by equations \ref{eq:rho} \& \ref{eq:csq}, it is straightforward to show that $Be(r_o) = -GM/R$ for $\gammae = \gamma$ (independent of both radius $r_o$ and $\gammae$). Thus for stellar envelopes with negligible mass it is possible to have $f \ll 1$ if $R \gg r_h$.   For more realistic massive stellar models, we find that $f \sim 0.1-1$ in the outer stellar envelope, with smaller values for more extended  envelopes (see Fig. \ref{fig:thermal} discussed in \S \ref{sec:applications}).

Equations \ref{eq:mdot}-\ref{eq:energy2} can be combined to yield the wind equation that describes the acceleration of the wind (again neglecting $L_{\rm rad}$):
\be
\frac{1}{v}\frac{dv}{dr}\left(v^2 - \frac{4}{3} c_s^2\right) = \frac{8}{3} \frac{c_s^2}{r} - \frac{\dot q}{3 v \rho} - \frac{GM}{r^2}.
\label{eq:wind}
\ee
Equation \ref{eq:wind} is of the usual form where the requirement that the flow smoothly evolve from subsonic to supersonic at a sonic point  implies that the left and right hand sides  of equation \ref{eq:wind} vanish so that\footnote{There are also breeze solutions that never go super-sonic.  These may be relevant to the early time dynamics in the numerical models presented in \S \ref{sec:numerics}, when the envelope mass just exterior to the heating region is so large that it stifles the wind.}
\be
{\rm Sonic \, Point} \ r_s \rightarrow  \ \ \  v^2 = \frac{4}{3}c_s^2 \ \ \ {\rm and} \ \ \ \frac{8}{3} \frac{c_s^2}{r_s} - \frac{\dot q}{3 v \rho} - \frac{GM}{r_s^2} = 0.
\label{eq:sonic}
\ee
Equations \ref{eq:sonic} are additional boundary conditions that specify the mass outflow rate $\dot M$ and the location of the sonic point $r_s$.   The solution of equation \ref{eq:sonic} depends on the relative value of two timescales at the sonic point, the heating timescale $t_{\rm heat} \sim \rho c_s^2/\dot q$ relative to the dynamical timescale $t_{\rm dyn} \sim r_s/c_s$.   We are working in the limit $t_{\rm heat} \gtrsim t_{\rm dyn}$, or else the heating would be dynamical and drive shocks.   In this limit the solution of equation \ref{eq:sonic} for the sonic point is the usual sonic point condition absent heating:
\be
c_s^2(r_s) \simeq \frac{3}{8}\frac{G M}{r_s}.
\ee

To solve for the wind mass loss rate $\dot M$ and sonic point radius $r_s$, we begin by noting that  because the sonic point is located at $r_s \gtrsim r_h$,  energy conservation implies $Be(r_s) \simeq Be(r \rightarrow \infty) \simeq v_\infty^2/2$, where $v_\infty$ is the asymptotic velocity of the flow at large radii.   Using equation \ref{eq:sonic}, $Be(r_s)$ can be shown to be $3GM/4r_s$, which implies
\be
r_s \simeq \frac{3}{2} \frac{G M}{v_\infty^2}.  
\label{eq:rs}
\ee
An expression for the asymptotic velocity $v_\infty$ can be obtained by combining equations \ref{eq:energyBe}  \& \ref{eq:Berh}:
\be
v_{\infty}^2 = \frac{2 \dot E}{\dot M} - f \, \ve^2(r_h).
\label{eq:vinfinity}
\ee

\begin{figure}
\resizebox{9cm}{!}{\plotone{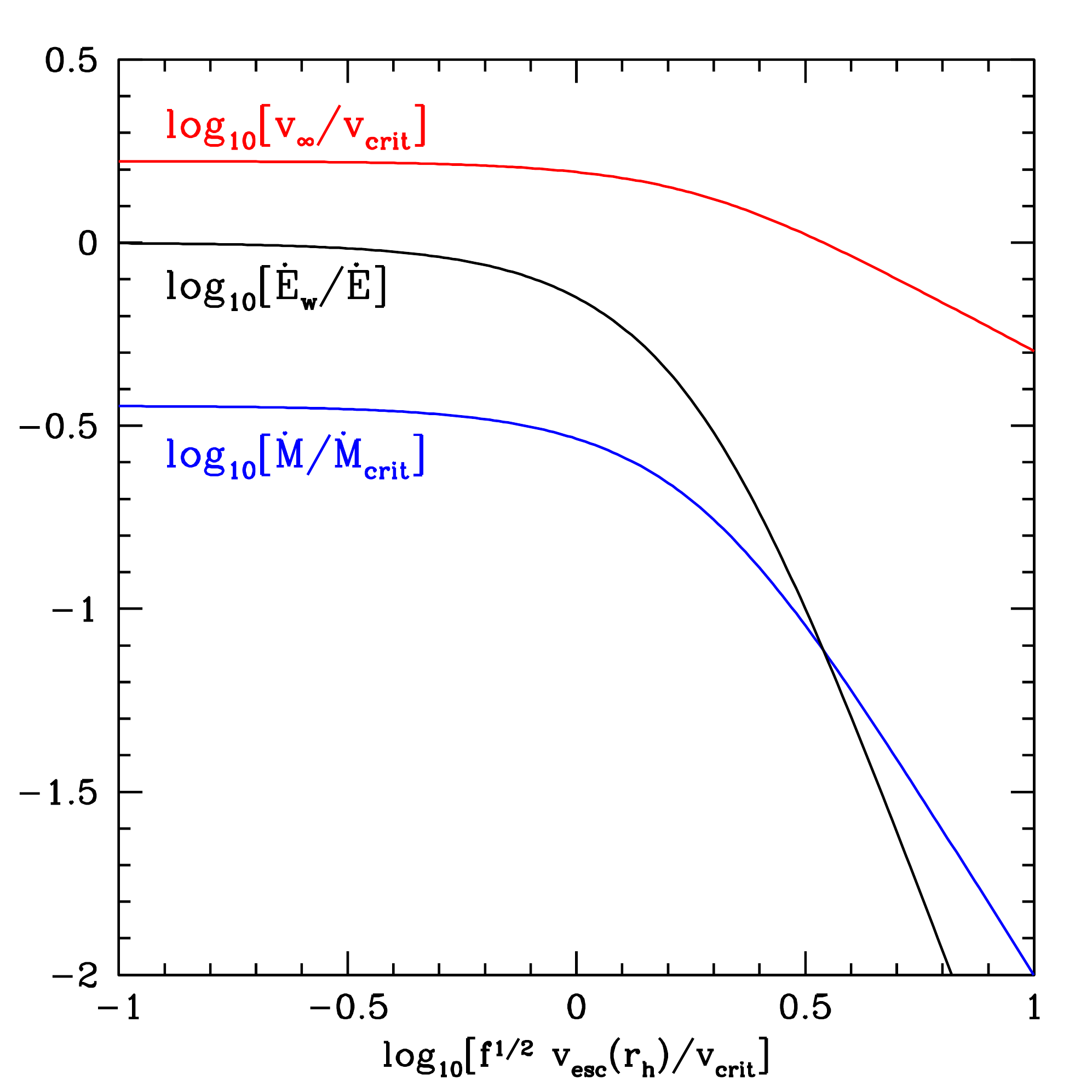}}
\caption{Analytic estimates of the properties of steady state spherically symmetric super-Eddington winds (see \S \ref{sec:analytics}): asymptotic wind speed $v_\infty$ (in units of $\vc$; see eq. \ref{eq:vcrit}), mass loss rate $\dot M$ (in units of $\dot M_{\rm crit}$; see eq. \ref{eq:mdotcrit}), and asymptotic wind  energy flux $\dot E_w$ (in units of the energy input rate $\dot E$).}
\label{fig:analytics}
\end{figure}

\begin{table*}
\centering
\begin{minipage}{17.5cm}
\caption{List of wind models evolved with FLASH, and summary of results\label{t:flash_models}\label{t:flash_results}.
Columns (left to right) show model name, heating rate, initial envelope radius, radial width
of the heating region, dimensionality, ratio $v_{\rm esc}(R)/v_{\rm crit}$ (eq.~\ref{eq:vcrit}), total simulated
time, time to achieve steady-state in the sonic point evolution (${\rm d}\ln{r_s}/{\rm d}\ln t = 10^{-2}$),
steady-state mass loss rate in absolute units and normalized to the critical value (eq.~\ref{eq:mdotcrit}),
ratio of steady-state outgoing wind energy loss rate to input power, and ratio of the asymptotic
sonic point radius to the initial envelope radius. For uniformity, the mass and energy loss in 2D and 3D
have been multiplied by $4\pi/\Delta \Omega$, where $\Delta \Omega$ is the solid angle subtended
by the computational domain. Results have been rounded to two significant digits.}
\begin{tabular}{lccccccccccc}
\hline
{Model}&
{$\dot{E}$} &
{$R$} &
{$\Delta r_h$} &
{Dim.} &
{$v_{\rm esc}(R)/v_{\rm crit}$} &
{$\Delta t_{\rm sim}$} &
{$\Delta t_{\rm steady}$} &
{$\dot{M}$} &
{$\dot{M}/\dot{M}_{\rm crit}$} &
{$\dot{E}_{\rm w}/\dot{E}$} &
{$r_{\rm s}/R$} \\
{} & {$({r_h}^2 \rho_h {\ve(r_h)}^3)$} & {$(r_h)$} & {$(r_h)$} &
     {} & & \multicolumn{2}{c}{$(r_h/\ve(r_h))$}  & {(${r_h}^2 \rho_h{\ve(r_h)}$)} & {} & {} & {} \\
\hline                                                                                 
L005R2.5-1d   & 5.2E-4 & 2.5 & 0.075 & 1D & 5.4     & 2.8E+4 & 1.4E+4 & 2.5E-3 & 3.3E-2 & 1.3E-2 & 50 \\ 
L015R2.5-1d   & 1.5E-3 &     &       &    & 4.3     & 8.8E+3 & 5.2E+3 & 7.4E-3 & 5.1E-2 & 2.9E-2 & 25 \\ 
L046R2.5-1d   & 4.6E-3 &     &       &    & 3.5     & 5.3E+3 & 1.6E+3 & 2.2E-2 & 7.9E-2 & 6.1E-2 & 12 \\ 
L150R2.5-1d   & 1.5E-2 &     &       &    & 2.7     &         & 4.1E+2 & 6.8E-2 & 1.2E-1 & 1.3E-1 & 5.0 \\ 
L150R5.0-1d   &        & 5   &       &    & 2.2     &         & 6.1E+2 & 1.2E-1 & 1.7E-1 & 1.8E-1 & 3.3 \\ 
L150R10-1d    &        & 10  &       &    & 1.6     &         & 5.5E+2 & 2.0E-1 & 2.4E-1 & 3.1E-1 & 1.5 \\ 
L150R25-1d    &        & 25  &       &    & 1.1     &         & 4.1E+2 & 2.9E-1 & 3.0E-1 & 5.2E-1 & 5.4E-1 \\ 
L150R50-1d    &        & 50  &       &    & 8.1E-1  &         & 2.3E+2 & 3.3E-1 & 3.2E-1 & 6.4E-1 & 2.5E-1 \\ 
L150R100-1d   &        & 100 &       &    & 6.0E-1  &         & 1.8E+1 & 3.6E-1 & 3.3E-1 & 7.2E-1 & 1.2E-1 \\ 
\noalign{\smallskip}
L005R2.5-2d   & 5.2E-4 & 2.5 & 0.075 & 2D & 5.4     & 2.8E+4 & 1.4E+4 & 2.6E-3 & 3.4E-2 & 1.4E-2 & 51  \\
L015R2.5-2d   & 1.5E-3 &     &       &    & 4.3     & 8.8E+3 & 3.8E+3 & 7.5E-3 & 5.2E-2 & 3.0E-2 & 24  \\
L046R2.5-2d   & 4.6E-3 &     &       &    & 3.5     & 5.3E+3 & 1.6E+3 & 2.2E-2 & 7.7E-2 & 6.2E-2 & 12  \\
L150R2.5-2d   & 1.5E-2 &     &       &    & 2.7     &         & 4.0E+2 & 6.7E-2 & 1.2E-1 & 1.3E-1 & 5.0 \\
L150R5.0-2d   &        & 5   &       &    & 2.2     &         & 6.1E+2 & 1.2E-1 & 1.7E-1 & 1.9E-1 & 3.2 \\
L150R10-2d    &        & 10  &       &    & 1.6     &         & 5.5E+2 & 2.0E-1 & 2.4E-1 & 3.1E-1 & 1.5 \\
L150R25-2d    &        & 25  &       &    & 1.1     &         & 4.0E+2 & 2.9E-1 & 3.0E-1 & 5.3E-1 & 5.3E-1 \\
L150R50-2d    &        & 50  &       &    & 8.1E-1  &         & 2.3E+2 & 3.4E-1 & 3.3E-1 & 6.4E-1 & 2.5E-1 \\
L150R100-2d   &        & 100 &       &    & 6.0E-1  &         & 1.4E+2 & 3.6E-1 & 3.3E-1 & 7.2E-1 & 1.2E-1 \\
\noalign{\smallskip}
L046R2.5-3d   & 4.6E-3 & 2.5 & 0.075 & 3D & 3.5     & 2.8E+3 & 1.8E+3 & 2.2E-2 & 7.8E-2 & 6.1E-2 & 12 \\
\hline
\end{tabular}
\end{minipage}
\end{table*}

The mass loss rate can be evaluated at the sonic point as
\be
\dot M =  4 \pi r_s^2 \rho(r_s) v(r_s) = \sqrt{27} \pi \, \frac{(GM)^2 \, \rho(r_s)}{v_\infty^3}
\label{eq:mdot1}
\ee
where we have used equations \ref{eq:sonic} and \ref{eq:rs}.   To simplify equation \ref{eq:mdot1} further, we require the density at the sonic point $\rho(r_s)$.    Since the flow is subsonic interior to the sonic point, this can be estimated to reasonable accuracy by solving hydrostatic equilibrium interior to the sonic point with a boundary condition at small radii that determines the normalization of the density.     To do so, we proceed as follows.  For radii satisfying $r_h \lesssim r \lesssim r_s$ there is no heating and so the outflow is adiabatic with $\gamma = 4/3$ by assumption.   Moreover the flow is subsonic and so roughly in hydrostatic equilibrium.   As a result, the density profile between the heating region and the sonic point is well described by equation \ref{eq:rho} with $\gammae \rightarrow 4/3$, $\Ke \rightarrow \Kw$, and $R \rightarrow \infty$ (the latter because the wind is unbound).   This implies
\be
\rho(r) \simeq \frac{1}{4^3 \Kw^3} \left(\frac{GM}{r}\right)^3  \hspace{1cm} r_h \lesssim r \lesssim r_s
\label{eq:rhowind}
\ee
The wind entropy $\Kw$ can be calculated by using pressure balance across the heating radius.   We define $r_h^-$ and $r_h^+$ to be radii just inside and outside the heating radius, respectively.   Pressure balance implies $P(r_h^-) \simeq P(r_h^+) = \Kw \rho(r_h^+)^{4/3} = (GM)^4/(4^4 \Kw^3 r_h^4)$.   Hydrostatic equilibrium in the stellar envelope requires $P(r_h^-) \simeq \rho_{\rm env}(r_h) GMH/r_h^2$ where $H$ is the local pressure scale height in the envelope near $r_h^-$.  We define the hydrostatic envelope mass as
\be
\Me \equiv 10 \times 4\pi \, r_h^2 \, H \, \rho_{\rm env}(r_h)
\label{eq:Me}
\ee
The factor of 10 in equation \ref{eq:Me} is arbitrary but is motivated by the fact that for a $\gammae = 4/3$ envelope $H \sim r/4$, $\rho_{\rm env} \propto r^{-3}$ and thus each decade in radius contributes comparably to the total mass.  For a spatially extended envelope the total mass is thus a multiple $\sim 10$ of the local mass at $r_h$.    Despite this particular motivation, equation \ref{eq:Me} can simply be viewed as a definition of $\Me$ used to set the density scale for the envelope mass and thus the outflow.     Note also that if the size of the heating region is comparable to or larger than the local scale-height $H$ then $P(r_h^-) \simeq P(r_h^+)$ is not a good approximation.   Since $P(r_h^-) \propto \Me$, this can be roughly captured in what follows by a  reduction in the envelope mass $\Me$.

With equation \ref{eq:Me}, pressure balance across the heating radius implies
\be
\Kw^3 \simeq \frac{10 \pi}{4^3} \frac{(GM)^3}{\Me}.
\label{eq:Kw}
\ee
Combining equations \ref{eq:rs}, \ref{eq:rhowind} and \ref{eq:Kw} we find the wind density at the sonic point $r_s$ of $\rho(r_s) \simeq \Me/10 \pi r_s^3$ and thus arrive at our final expression for the mass outflow rate
\be
\dot M \simeq \frac{4}{15 \sqrt{3}} \frac{\Me}{M} \frac{v_\infty^3}{G} =  \frac{4}{15 \sqrt{3}} \frac{\Me}{G M} \left(\frac{2 \dot E}{\dot M} - f \ve^2(r_h)\right)^{3/2}.
\label{eq:mdot-final}
\ee
We reiterate that $\ve(r_h)$ here is the escape velocity of the stellar envelope just interior to the heating radius.  This is important because it sets the Bernoulli parameter of the stellar envelope and thus the energy that must be supplied to unbind mass from the star.     

The solution of equation \ref{eq:mdot-final} depends on the relative magnitude of the  escape velocity $\ve(r_h)$ to a characteristic velocity of the problem $\vc$, where
\be
\ve(r_h) \simeq 620 \, \kms \, M^{1/2}_{30} \, r_{h,30}^{-1/2},
\label{eq:ve}
\ee
$M = 30 \, M_{30} \,M_\odot$, $r_h = 30 \, r_{h,30} \, R_\odot$, and
\be
v_{\rm crit} \equiv \left(\frac{M}{\Me} \, G \, \dot E\right)^{1/5} \simeq 190 \kms \, \left(\frac{M}{10^3 \, \Me}\right)^{1/5} \, \dot E_7^{1/5}
\label{eq:vcrit}
\ee
where $\dot E_7 = \dot E/(10^7 L_\odot)$; note that $10^7 L_\odot$ corresponds to about 10 times the electron-scattering Eddington luminosity for our fiducial $30 M_\odot$ star.   In equation \ref{eq:vcrit} and what follows we take a typical $\Me \sim 10^{-3} M$.   As discussed in \S \ref{sec:applications} (Fig. \ref{fig:thermal}), this is required in order to explain the timescale of observed super-Eddington mass loss from massive stars.

We also define a characteristic mass loss rate using
\be
\dot E = \frac{1}{2} \, \dot M_{\rm crit} \vc^2 \rightarrow \dot M_{\rm crit} \equiv \frac{2 \vc^3 \Me}{G M} 
\label{eq:mdotcrit}
\ee
in which case equation \ref{eq:mdot-final} can be written in dimensionless form as
\be
\left(\frac{\dot M}{\dot M_{\rm crit}}\right)^5 = \frac{4}{675}\left(1-\frac{\dot M}{\dot M_{\rm crit}} \frac{f \, \ve(r_h)^2}{\vc^2}\right)^3.
\label{eq:mdot-dimensionless}
\ee

Figure \ref{fig:analytics} shows the numerical solution of equations \ref{eq:vinfinity} and \ref{eq:mdot-dimensionless} for the wind asymptotic velocity $v_\infty$ in units of $\vc$, the mass outflow rate in units of $\dot M_{\rm crit}$ and the asymptotic wind kinetic energy flux $\dot E_w = 0.5 \dot M v_\infty^2$ in units of the energy input rate $\dot E$.    These results show that there are two regimes with different wind physics:
$$\boxed{\rm Regime \, 1:  \vc \gtrsim f^{1/2} \ve(r_h)}$$
This corresponds to relatively large stellar progenitors and/or low mass, weakly bound envelopes.   In this case $\dot E/\dot M \gg f \ve^2(r_h)$ and the solution to equations \ref{eq:vinfinity} and \ref{eq:mdot-final} is $\dot M \simeq 0.36 \, \dot M_{\rm crit}$, i.e.,
\be
\dot M \simeq 0.72 \left(\frac{\Me}{M G}\right)^{2/5}  \dot E^{3/5} \simeq 1 \mspy \, \left(\frac{M}{10^{3} \, \Me}\right)^{-2/5} \, \dot E_7^{3/5}
\label{eq:Mdot-vclarge}
\ee
and
\be
v_\infty \simeq 1.7 \vc \simeq 300 \kms \, \left(\frac{M}{10^3 \, \Me}\right)^{1/5} \, \dot E_7^{1/5}.
\label{eq:v-vclarge}
\ee
In this regime the sonic point is located at 
\be
r_s \simeq 0.52 \, (GM)^{3/5} \Me^{2/5} \, \dot E^{-2/5} \simeq 80 \, R_\odot \, M_{30} \left(\frac{M}{10^{3} \, \Me}\right)^{-2/5} \, \dot E_7^{-2/5}
\label{eq:rs-vclarge}
\ee
and the asymptotic wind kinetic power is given by
\be
\dot E_w = \frac{1}{2} \dot M v_\infty^2 \simeq \dot E.
\label{eq:Ew-vclarge}
\ee
Equation \ref{eq:Ew-vclarge} corresponds to nearly all of the energy deposited in the stellar envelope going into driving a wind.   

Note that equation \ref{eq:rs-vclarge} for the sonic point location can also be rewritten as
\be
r_s \simeq \frac{r_h}{4} \, \left(\frac{\ve(r_h)}{\vc}\right)^2.
\label{eq:rs-rh}
\ee
Physically, solutions must have $r_s \gtrsim r_h$ since the sonic point cannot lie interior to the heating radius.   This implies that we require $\ve(r_h) \gtrsim 2 \vc$ and thus $f \lesssim1$ for physical wind solutions to both be in the regime $\vc \gtrsim f^{1/2} \ve(r_h)$ and have $r_s \gtrsim r_h$.   Note that $\ve(r_h) \gtrsim 2 \vc$ and equation \ref{eq:v-vclarge} together imply that the maximal value of the asymptotic wind speed is $\sim \ve(r_h)$.  

What happens if  $\vc \gtrsim 0.5 \ve(r_h)$ (and hence eq. \ref{eq:rs-rh} implies $r_s \lesssim r_h$)?  This is certainly physically realizable.    A simple estimate shows that $t_{\rm thermal}(r_h)/t_{\rm dyn}(r_h) \sim (\ve(r_h)/\vc)^5$ so that $\vc \gtrsim \ve(r_h)$ corresponds to heating on less than or of order a dynamical time, which will lead to a strong shock rather than a steady wind.   
$$ \boxed{\rm Regime \, 2: f^{1/2} \ve(r_h) \gtrsim \vc}$$
This corresponds to relatively compact stellar progenitors and/or higher mass, tightly bound  envelopes.   In this case the solution to equation \ref{eq:mdot-final} satisfies $\dot E/\dot M \gg v_\infty^2$ and
\be
\dot E \simeq  \frac{f}{2} \dot M \ve^2(r_h) \rightarrow \dot M \simeq \frac{\dot E \, r_h}{G \, M f} \simeq  0.3 \, \mspy \,  \dot E_7 \, M_{30}^{-1} \, r_{h,30} \, f^{-1}
\label{eq:Mdot-vcsmall}
\ee
and 
\be
v_\infty \simeq 1.9 \left(\frac{\dot E r_h}{f \Me}\right)^{1/3} \simeq 200 \, \kms \left(\frac{\dot E_7 r_{h,30}}{f M_{30}}\right)^{1/3} \left(\frac{M}{10^3 \Me}\right)^{1/3}.
\label{eq:v-vcsmall}
\ee
In this regime the sonic point is located at $r_s \simeq 0.14 r_h f^{2/3} (\ve(r_h)/\vc)^{10/3}$, i.e.,
\be
r_s \simeq 200 \, R_\odot \  \dot E_7^{-2/3} M_{30}^{5/3} r_{h,30}^{-2/3} f^{2/3} \left(\frac{M}{10^3 \Me}\right)^{-2/3}
\label{eq:rs-vcsmall}
\ee
and the asymptotic wind energy flux is given by
\be
\frac{\dot E_w}{\dot E} \simeq \frac{v_{\infty}^2}{f \ve(r_h)^2} \simeq 0.1 \, \dot E_7^{2/3} M_{30}^{-5/3} r_{h,30}^{5/3} \, f^{-5/3} \left(\frac{M}{10^3 \Me}\right)^{2/3}.
\label{eq:Ew-vcsmall}
\ee
In this regime the asymptotic energy flux of the wind is  small compared to the energy supplied to the stellar envelope.  Most of the input energy is used to unbind the gas from the potential of the star, as implied by $\dot E \simeq |Be(r_o)| \dot M \simeq 0.5 f \dot M \ve^2(r_h)$ in equation \ref{eq:Mdot-vcsmall}.
\vspace{-0.3cm}
\subsection{Validity of the Adiabatic Approximation}
\label{sec:adiabatic}
In deriving the wind properties in this section we have neglected photon transport of energy and assumed that the outflow is adiabatic.   This requires that $t_{\rm diff} \simeq r^2 \kappa \rho/c \gtrsim t_{\rm exp} \simeq r/v$ at least out to the sonic point ($\kappa$ here is the opacity).   Using equations \ref{eq:sonic}, \ref{eq:rs}, \& \ref{eq:mdot1} it is straightforward to show that the condition for adiabaticity at the sonic point takes the intuitive form $\dot E_w \gtrsim L_{\rm Edd}$ (this also follows directly from equation \ref{eq:energy2} by noting that $\dot M Be = \dot E_w$).    Physically, this condition states that if the asymptotic kinetic power of the wind is greater than the Eddington luminosity, the photons are trapped in the wind at the radii where the wind properties are set (between the heating region and the sonic point).  The wind properties can thus be calculated neglecting radiation diffusion and assuming a $\gamma = 4/3$ radiation dominated fluid.   Since $\dot E_w/\dot E$ declines with increasing $\ve(r_h)/\vc$ (Fig. \ref{fig:analytics} \& eq. \ref{eq:Ew-vcsmall}), in practice the current solutions formally apply only if $\ve(r_h)/\vc \lesssim 1.7 f^{-1/2} (\dot E/L_{\rm Edd})^{3/10}$.       

\vspace{-0.3cm}
\subsection{Reformulation in Terms of an Eddington Ratio $\Gamma$}
\label{sec:Gamma}

Using the diffusion equation to relate the radiation flux $F_{\rm rad}$ and the radiation pressure gradient, the momentum equation (eq. \ref{eq:mom}) can be rewritten as
\be
v \frac{dv}{dr} = \frac{\kappa}{c} \left(F_{\rm rad} - F_{\rm Edd}\right) = (\Gamma - 1) \frac{GM}{r^2}
\label{eq:mom-gamma}
\ee
where $F_{\rm Edd} = cg/\kappa$, $g$ is the gravitational acceleration, $\Gamma = F_{\rm rad}/F_{\rm Edd}$, and we have neglected gas pressure.   

Equation \ref{eq:mom-gamma} shows that it is possible, of course, to reformulate the results in this section in terms of an effective Eddington ratio $\Gamma$.   We believe, however, that this somewhat obscures the underlying physics.  In particular, the model developed in this section is most appropriate precisely when the  radiation flux is small compared to the energy flux in the wind; $F_{\rm rad}$ and hence $\Gamma$ are thus not the key dynamical variables.  Moreover, we show in \S \ref{sec:flash} that numerical simulations without any radiation diffusion reproduce well the analytic models developed here; all that is required is a fluid with a radiation ($\gamma = 4/3$) equation of state.  And we show in \S \ref{sec:mesa} that inclusion of radiation diffusion does not significantly change the properties of the solutions when the constraint $\dot E_w \gtrsim L_{\rm Edd}$ of \S \ref{sec:adiabatic} is satisfied.    The key point is that when the optical depths are sufficiently high, photons are trapped, the outflow is adiabatic, and the wind energy is roughly conserved outside the heating region.  In our models, the outflow properties in this limit are insensitive to the exact opacity because once the photons are trapped out to the sonic point, it doesn't matter how trapped they are.    

\vspace{-0.4cm}
\section{Numerical Wind Solutions}
\label{sec:numerics}

In this section we present time-dependent numerical hydrodynamic models  that
further elucidate the physics of super-Eddington winds.  These demonstrate
good agreement with the analytic solution in \S \ref{sec:analytics}.   We
present two sets of numerical models.  The first (\S \ref{sec:flash}) are one-
(1D), two- (2D), and three dimensional (3D) simulations using the Eulerian
hydrodynamics code FLASH of the model problem solved analytically in \S
\ref{sec:analytics}:  a $\gamma = 4/3$ fluid subjected to external energy
input.   The principal contribution of these FLASH simulations is that they
demonstrate that multi-dimensional effects in general, and convection in
particular, do not noticeably change the properties of the wind solution.  The
second numerical models we present are 1D hydrodynamic simulations with MESA
(\S \ref{sec:mesa}).  The key features of this simulation are that it includes
radiation diffusion as well as a realistic stellar progenitor.

\vspace{-0.4cm}
\subsection{Time-Dependent Hydrodynamics with FLASH}
\label{sec:flash}

In presenting our models evolved with FLASH, we focus on comparing
the global wind quantities with the analytic predictions in steady-state,
and on the potential role of multidimensional effects.

\vspace{-0.3cm}
\subsubsection{Problem Setup and Models Evolved}

\begin{figure*}
\includegraphics*[width=\columnwidth]{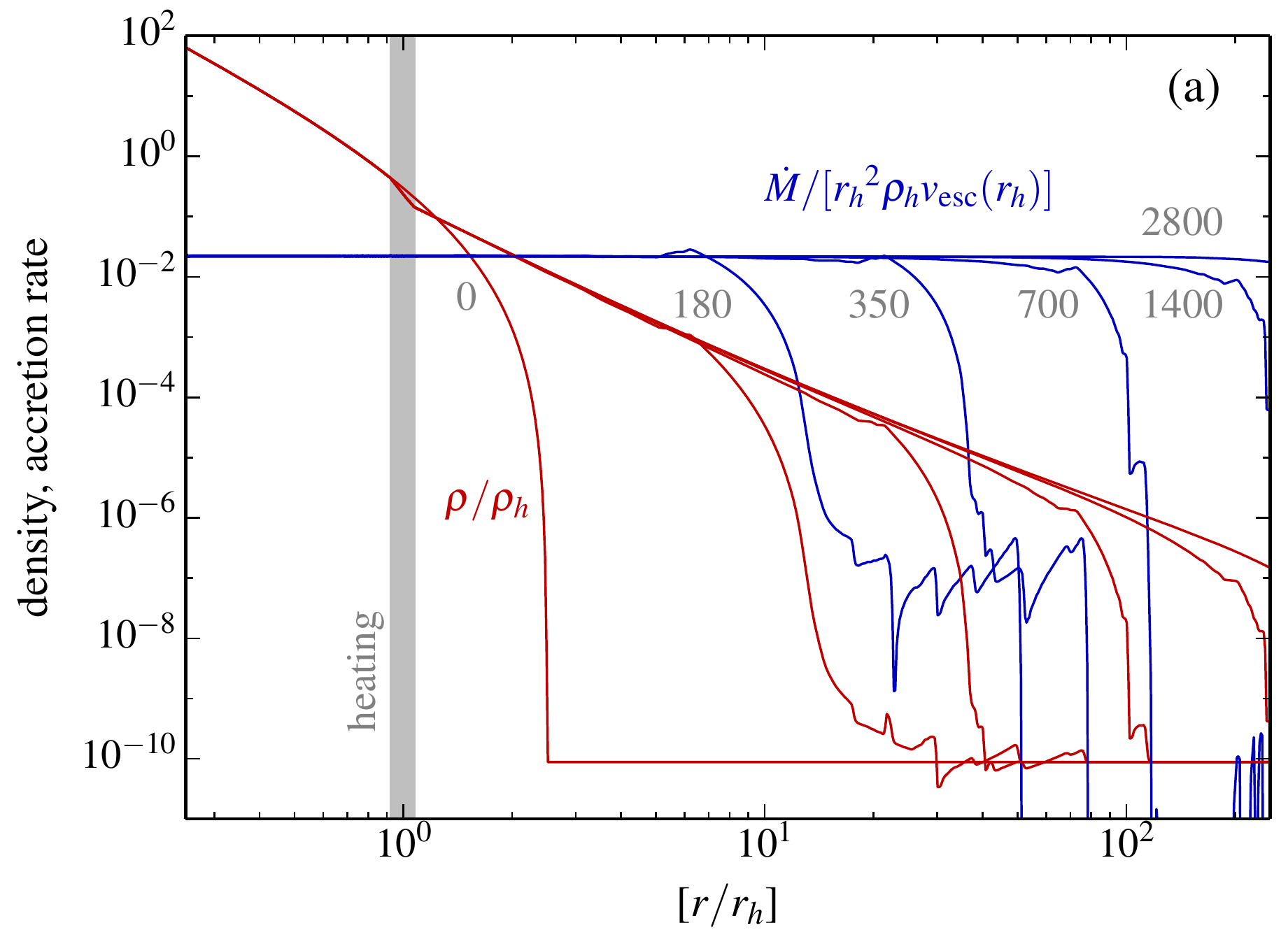}
\includegraphics*[width=0.99\columnwidth]{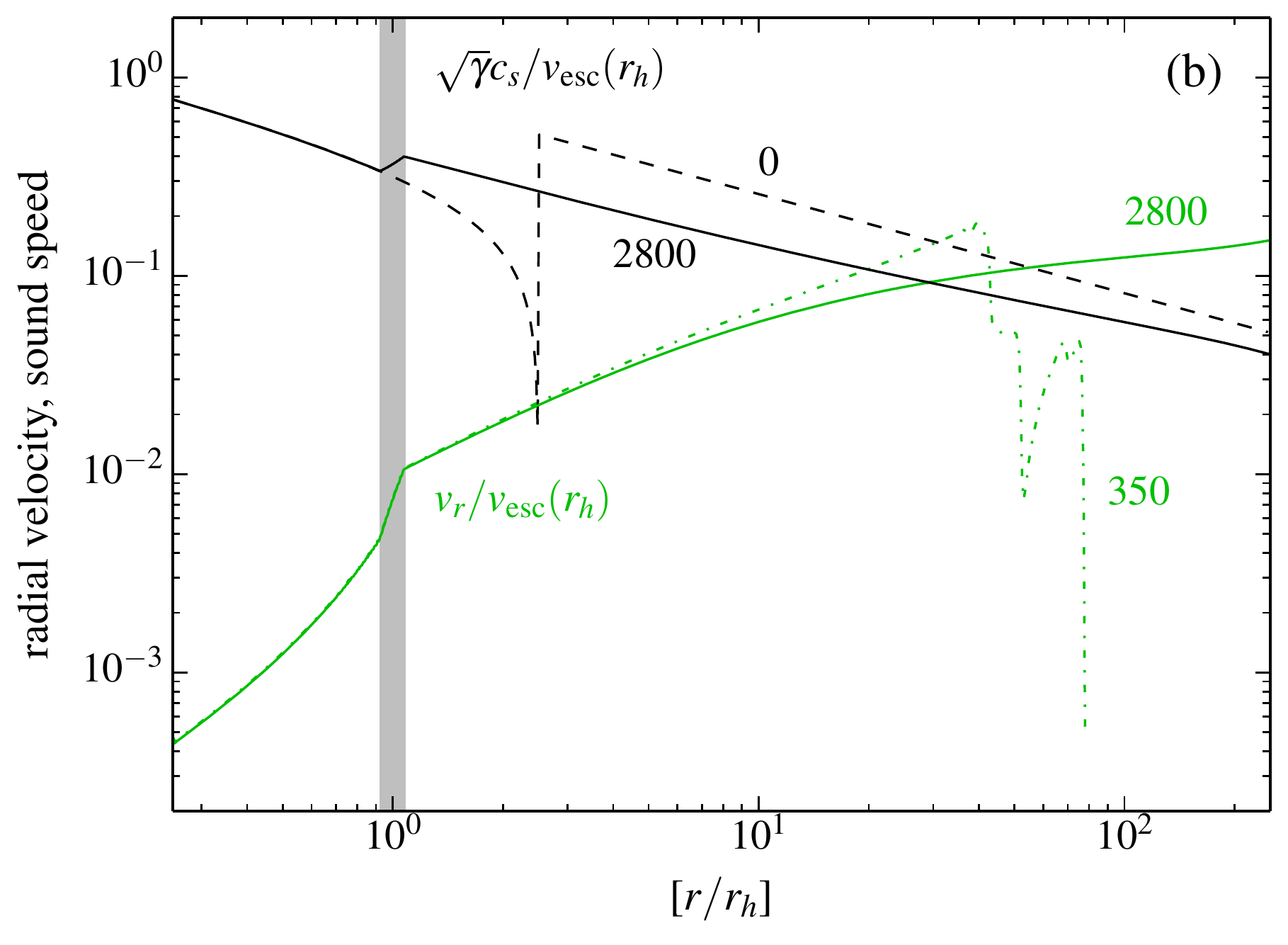}
\caption{Evolution of the spherically symmetric wind model L046R2.5-1d with FLASH, illustrating the transition
to steady-state. Panel (a) shows profiles of density (red) and mass loss
rate (blue) at the labeled times 
(in units of $r_h/\ve(r_h)$, rounded to two significant digits). 
The gray shaded
region denotes the radial range where heating is imposed. Panel (b) shows sound
speed (black) and radial velocity (green) at the labeled times.}
\label{f:flash_1d_snapshots}
\end{figure*}

\begin{figure}
\includegraphics*[width=\columnwidth]{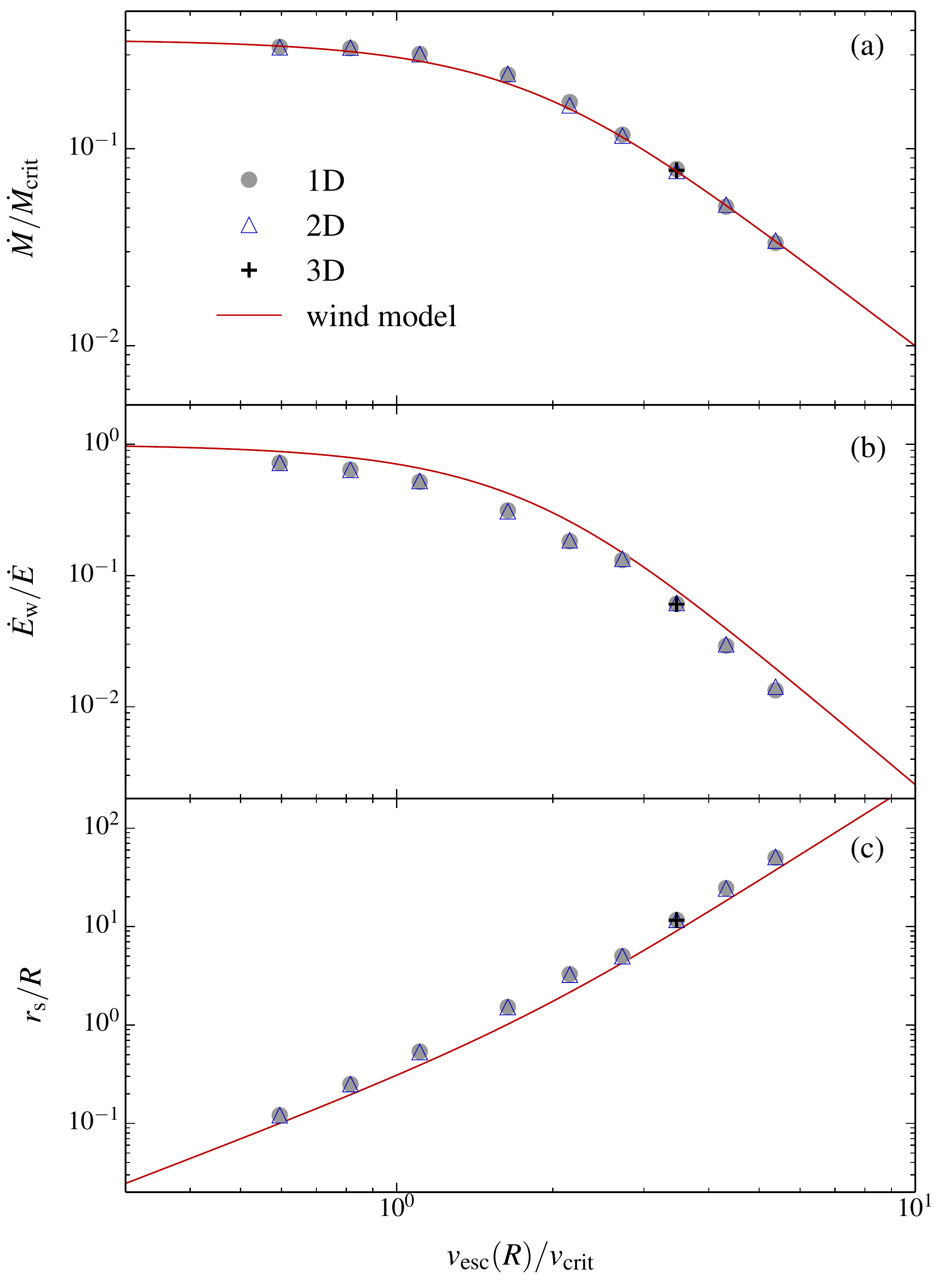}
\caption{Comparison between steady-state quantities from FLASH models (Table~\ref{t:flash_models},
gray circles, blue triangles, and black crosses) and the 
analytic wind model (red curves; eq.~\ref{eq:mdot-dimensionless} and 
Fig.~\ref{fig:analytics}).   There is essentially no difference between the 1D and 
multi-dimensional FLASH models.   \emph{Top:} Mass loss rate normalized by its critical 
value (eq.~\ref{eq:mdotcrit}). \emph{Middle:} Wind
energy loss rate (eq.~\ref{eq:energy_loss_flash}) normalized by the input heating rate.  \emph{Bottom:} 
Ratio of sonic point radius $r_s$ to initial envelope radius $R$.}
\label{f:flash_wind_steady}
\end{figure}

We use FLASH3 \citep{fryxell00,dubey2009} to
solve the equations of mass, momentum, and energy conservation, subject to
the gravity of a point mass and additional energy input.   
The public version of the code has been modified to allow for a
grid of non-uniform spacing in spherical polar coordinates \citep{F12,F15}.
The equation of state is that of an ideal gas with
adiabatic index $\gamma=4/3$.   No radiation transfer is included.   A constant heating
rate per unit volume is applied in the radius range $[r_h -\Delta r_h,r_h +\Delta r_h]$.   The FLASH models thus solve the identical problem posed in \S \ref{sec:analytics}, but include time dependence and multi-dimensional effects.   

We initialize the analytic profile in equation~(\ref{eq:rho}), with vanishing velocities.
We adopt a unit system based on the radius $r_h$, the initial density $\rho_h$, and 
escape speed $\ve(r_h)$
at the location where heating is applied.
The problem is determined by three dimensionless numbers: the initial envelope
radius $R/r_h$, the energy deposition rate 
$\dot{E} / ({r_h}^2 \rho_h {\ve(r_h)}^3)$
and the width of the heating region $2\Delta r_h/r_h$.
For $r \geq R$, the domain is initially filled with an ambient medium of constant density
$\rho_{\rm amb}\lesssim 10^{-10}\rho_h$, so that the mass in this ambient medium is
negligible compared to that in the envelope and subsequent wind. The initial
pressure in the ambient medium is $p_{\rm amb} = GM\rho_{\rm amb}/r$.   
The details of the initial ambient medium at $r \geq R$ are irrelevant to the subsequent evolution.

\begin{figure*}
\includegraphics*[width=\textwidth]{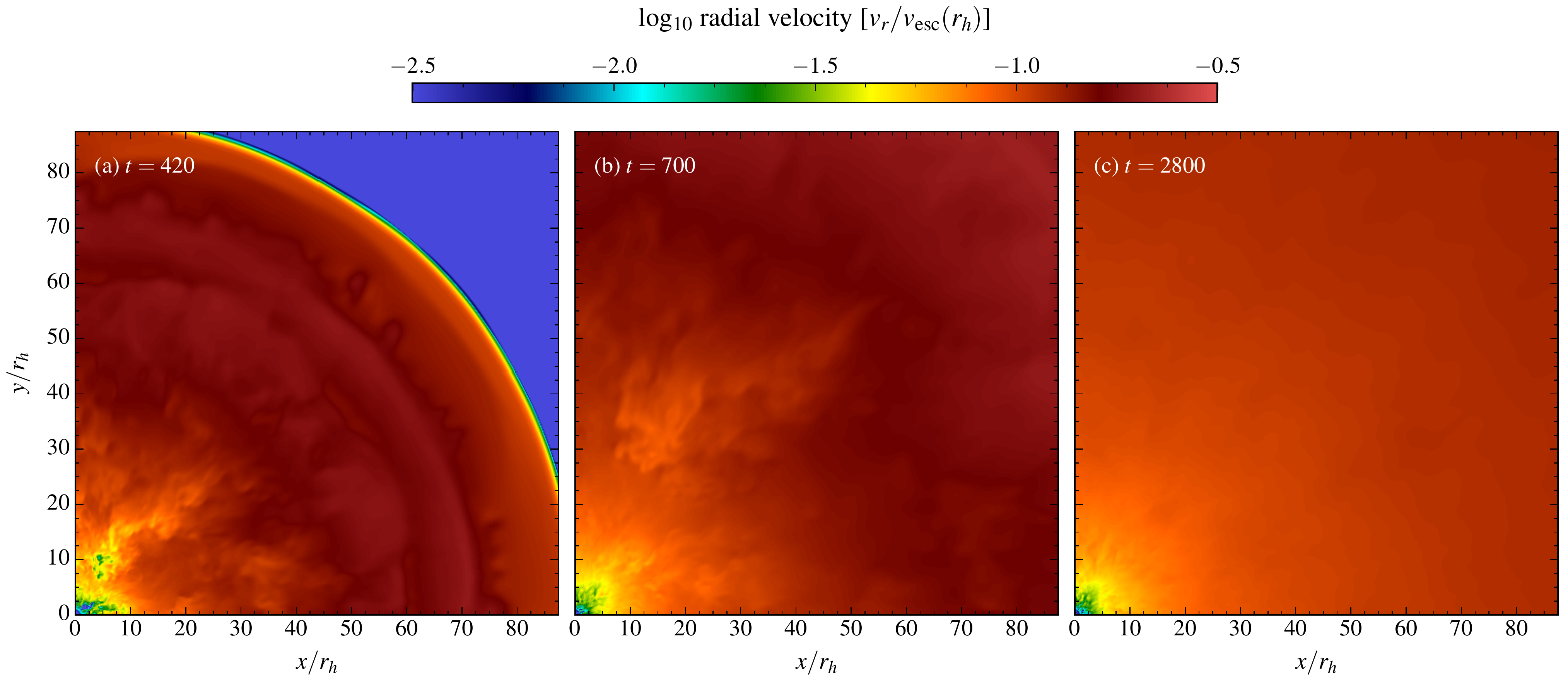}
\caption{Radial velocity on the equatorial plane at different times 
in the evolution of the 3D FLASH simulation L046R2.5-3d. 
At early times convection is important at small radii, and the expansion of the wind is not completely
spherical.  At later times, however, when the wind has reached steady state, the velocity field at large radii
is smooth and 
mostly
spherically symmetric. Times are labeled in units of $r_h/\ve(r_h)$, rounded to two significant digits.}
\label{f:flash_2d_snapshots}
\end{figure*}

The computational domain extends from $r=0.25r_h$ to $r=250r_h$, with logarithmic
radial spacing. In 2D and 3D,
the polar angle $\theta$ spans the range $[45^\circ,135^\circ]$ with uniform
spacing. In 3D the azimuthal angle $\phi$ spans the range $[0,90^\circ]$,
also with uniform spacing. The baseline resolution is such that
$\Delta r /r \simeq \Delta \theta = \Delta \phi \simeq 0.5^\circ \simeq 0.01$~rad.
The inner radial boundary condition is such that the ghost cells are filled
with the continuation of the solution implied by equation~(\ref{eq:rho}). This
forces the system to achieve steady state. The outer radial boundary condition
is set to outflow: zero gradient in all variables except the velocity, which
is set to zero if negative in the last active cell, or proportional to
$r^{-2}$ if positive, ensuring constant mass flux in the ghost cells.
In 2D and 3D, the angular boundary conditions in $\{\theta,\phi\}$ are periodic.

A key property of the FLASH models is that the gas in the polytropic atmosphere
and wind is not self-gravitating.  As a result, as noted below equation
\ref{eq:Berh}, the Bernoulli parameter of the  matter interior to the heating
radius is given by $Be = -GM/R$.    Note that this remains true even after the
wind has reached steady state and the matter that was at radius $r \sim R$ has
been unbound.   The reason is that the density, pressure, and sound speed
interior to the heating region do not change significantly even after the onset
of the wind; hence $Be$ does not change either.   As a result, the critical
dimensionless number that determines the properties of the wind is $v_{\rm
esc}(R)/v_{\rm crit}$.   This amounts to taking $f = r_h/R$ in the analytic
model in \S \ref{sec:analytics}, so that $f^{1/2}v_{\rm esc}(r_h) = v_{\rm
esc}(R)$.

Table~\ref{t:flash_models} shows the models we evolved.  We vary the
 heating rate or  initial envelope radius to obtain a range
of values of the ratio $v_{\rm esc}(R)/v_{\rm crit}(\dot{E},M_{\rm env})$.
Most models are run in 1D and 2D to study the effects of convection. One
model is run in 3D to quantify possible differences introduced by
the detailed properties of the turbulence in the transition to steady-state.
In most cases, the half-width of the heating region is $\Delta r_h = 0.075r_h$; a
comparison model with twice the width yields identical results, hence this choice 
is not important.

\vspace{-0.3cm}
\subsubsection{Overview of Evolution}

Figure~\ref{f:flash_1d_snapshots} shows the evolution of the density and mass
loss rate (Fig. \ref{f:flash_1d_snapshots}a) and velocity and sound speed (Fig.
\ref{f:flash_1d_snapshots}b) in a typical spherically-symmetric wind model.
The envelope begins to expand at radii $r > r_h$ on a thermal timescale (eq.
\ref{eq:thermal}). The increase in the entropy caused by the injection of heat
leads to a decrease in the density and an increase in the sound speed around
the heating region. As the system evolves, a power-law density
profile is established outside the heating radius.  This is qualitatively similar to the MESA model
shown in Figure~\ref{fig:density}, but in the present case the power-law
density profile is due to a wind rather than an extended convective envelope.
Figure~\ref{f:flash_1d_snapshots} shows that the mass loss rate adjusts to a
constant value as the envelope continues to expand, until all of the
computational domain reaches
steady-state.   The specific model in Figure \ref{f:flash_1d_snapshots} (L046R2.5-1d) has a
relatively high heating rate and thus at early times a shock forms at the outer edge of the
expanding envelope.   In the steady state wind there is no shock.   In addition, for lower heating rates, the expansion is fully subsonic without a shock.

The numerical solutions eventually reach a nearly perfect steady-state
throughout the computational volume.  Table~\ref{t:flash_models} shows the time
needed to achieve steady-state in the position of the sonic point $r_s$, which
we define as the first instant at which ${\rm d}\ln{r_s}/{\rm d}\ln t =10^{-2}$.
This time compares favorably with the thermal time estimated in equation~(\ref{eq:thermal}).
For example, the thermal time and time to steady-state for model L005R2.5-1d are 
$t_{\rm th}\simeq 5.5\times10^3\, r_h/\ve(r_h)$ and 
$\Delta t_{\rm steady} \simeq 1.4\times 10^4\,r_h/\ve(r_h)$, respectively.

\begin{figure}
\includegraphics*[width=\columnwidth]{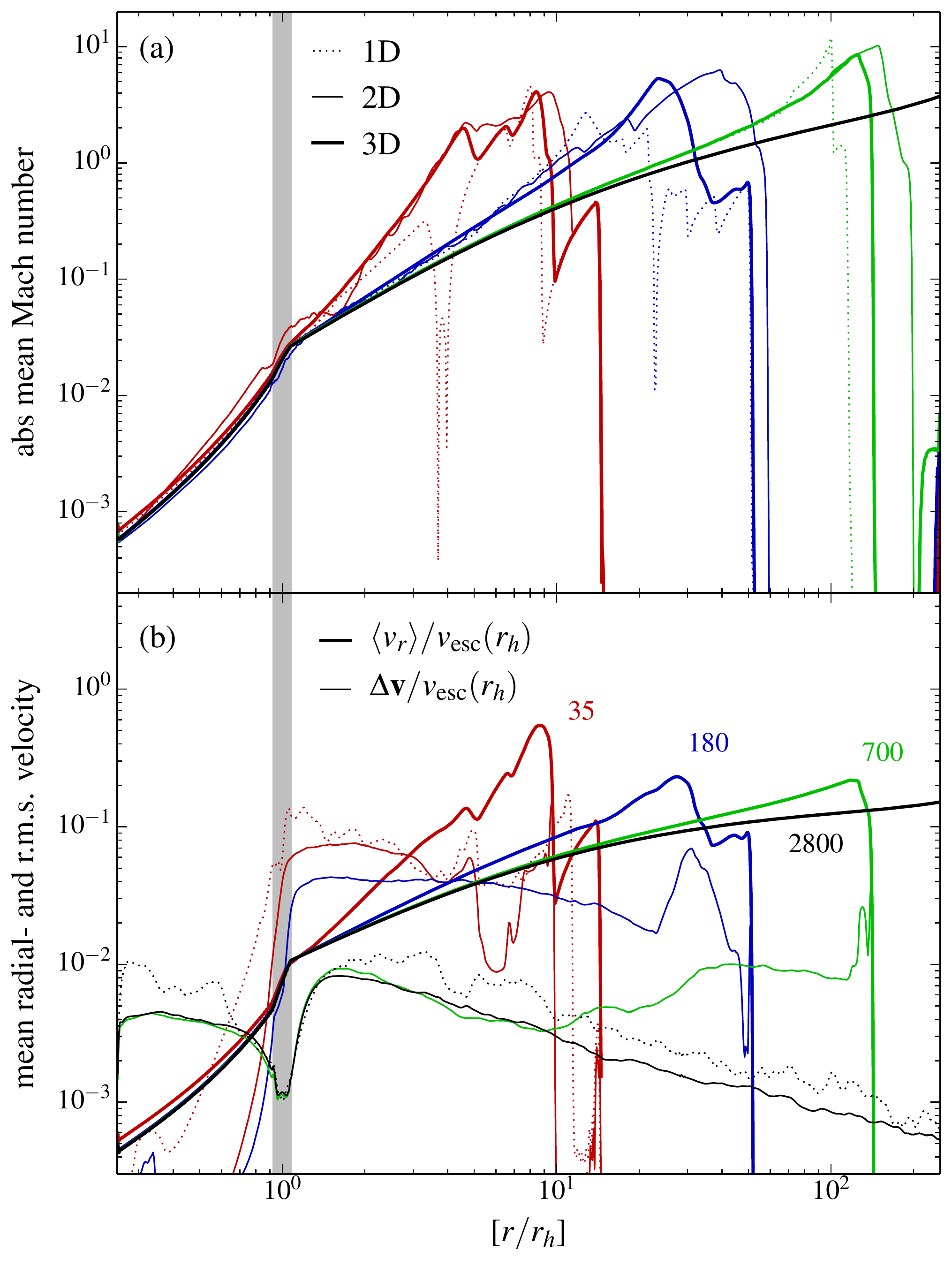}
\caption{Comparison between the velocity fields of 1D, 2D, and 3D wind models with FLASH.
\emph{Top:} Profiles of absolute value of the mean Mach number 
(eq.~\ref{eq:mean_mach_definition}) for model L046R2.5-3d and its 1D and 2D 
equivalent models (L046R2.5-1d and L046R2.5-2d, respectively).  There are modest
differences at large radii and early times, but in the steady state wind phase
the velocity profiles are remarkably similar in all cases.
\emph{Bottom:} Profiles of mean radial velocity and r.m.s. 
turbulent velocity (defined as in 
eqns.~\ref{eq:mean_definition}-\ref{eq:rms_definition}) 
for model L046R2.5-3d at different times (as
labeled in units of 
$r_h/\ve(r_h)$, rounded to two significant digits). The dotted lines show
the turbulent velocity field in 2D at $t=35$ and $2800$.
Note that the turbulent (convective) motions become unimportant at 
late times once a steady wind is established.}
\label{f:mach_flash_snapshots}
\end{figure}

\subsubsection{Steady-State Wind Properties \& Comparison to  Analytics}

The models shown in Table~\ref{t:flash_models} cover a factor $\sim 10$
in the ratio $v_{\rm esc}(R) / v_{\rm crit}$ and straddle the point
$v_{\rm esc}(R) = v_{\rm crit}$, which marks a qualitative transition
in the wind model of equation~(\ref{eq:mdot-dimensionless}). We compare three quantities
to the analytic predictions: the normalized mass loss rate $\dot{M}/\dot{M}_{\rm crit}$,
with $\dot M_{\rm crit}$ given by equation~(\ref{eq:mdotcrit}),
the ratio of wind power to input power $\dot{E}_w/\dot{E}$,
with
\begin{equation}
\label{eq:energy_loss_flash}
\dot E_{\rm w} = \int {\rm d}\Omega\, r^2 \rho v_r\, \left[\frac{1}{2}\mathbf{v}^2 + \frac{\gamma}{\gamma-1}\frac{p}{\rho}
            -\frac{GM}{r}\right]
\end{equation}
evaluated at some $r \gg r_h$, and the ratio of the sonic point to the initial envelope radius $r_s/R$, which is
a proxy for $v_\infty^2$ via equation~(\ref{eq:rs}). In the analytic solution, these three properties
are a function of the ratio $v_{\rm esc}(R) / v_{\rm crit}$ only.
In calculating $\dot M_{\rm crit}$ and $\vc$ for the numerical models, we  use the exact value of the initial
envelope mass (instead of the approximate definition in eq.~\ref{eq:Me}),
\begin{equation}
\label{eq:Menv_flash}
M_{\rm env}^{(num)} = 4\pi {r_h}^3 \rho_h \ln{(R/r_h)}.
\end{equation}
Table~\ref{t:flash_models} also reports the mass loss rate in our numerical units
$({r_h}^2\rho_h \ve(r_h))$, which are independent of  the definition of $M_{\rm env}$.

Figure~\ref{f:flash_wind_steady} compares the steady state wind properties from
the 1D, 2D, and 3D FLASH solutions to the analytic model.   There is good
overall agreement  in both the functional form and normalization of the
results, suggesting that equation~(\ref{eq:mdot-dimensionless}) is a reasonably
good description of the system dynamics.
Note that the extent of the quantitative agreement is somewhat sensitive to the
definition of $M_{\rm env}$. The value adopted in
equation~(\ref{eq:Menv_flash}) has been used consistently and is the most
accurate, although other definitions could potentially be used.

It is evident from Figure~\ref{f:flash_wind_steady} and Table~\ref{t:flash_models} that
the 1D, 2D, and 3D results are nearly identical. This suggests that multidimensional effects
in general (and convection in particular) have no significant role in maintaining
the wind once it has reached steady-state. This result is consistent with the
argument given in \S\ref{sec:conv-analytics}.
\vspace{-0.3cm}
\subsubsection{Multidimensional effects}

While multidimensional flows appear unimportant in maintaining
the steady wind, they can make a quantitative
difference in the transition to steady-state.
Figure~\ref{f:flash_2d_snapshots} illustrates the overall
magnitude of multidimensional flows,
showing snapshots of the radial velocity on the equatorial plane 
at a few instants in the
evolution of the 3D model L046R2.5-3d. Initially, convection is important
at small radii, and the expansion of the wind is not completely
spherical.
At later times, however, the velocity field at large radii
becomes increasingly smooth, with 
significant
non-sphericities maintained
only near the heating region.

We quantify the effects of non-spherical flows by taking angular averages of velocities.
The mean and r.m.s fluctuation of a quantity $A$ at a given radius are 
\begin{eqnarray}
\label{eq:mean_definition}
\langle A \rangle & = & \frac{\int{\rm d}\Omega \rho A \,\,}{\int {\rm d}\Omega \rho}\\
\label{eq:rms_definition}
\Delta A      & = & \left(\langle A^2\rangle - \langle A\rangle^2\right)^{1/2}.
\end{eqnarray}
Figure~\ref{f:mach_flash_snapshots}a shows profiles of the absolute value of the
mean radial Mach number
\begin{equation}
\label{eq:mean_mach_definition}
\mathcal{M} \equiv \frac{|\langle v_r\rangle|}{\sqrt{\gamma \langle c_s^2\rangle}}
\end{equation}
for model L046R2.5-2d, compared with its 1D and 2D equivalents. At early times, the
leading edge of the expanding wind lies at the same location in all
cases, while later the 2D model evolves faster than 3D and 1D. Nevertheless, the
final Mach number profiles are indistinguishable once the system has reached steady-state.

Figure~\ref{f:mach_flash_snapshots}b shows the mean radial velocity
and the r.m.s fluctuation of the total velocity (eq.~\ref{eq:rms_definition}), 
providing further insight into the magnitude of multidimensional flows. At
early times, convective motions are strongest in regions immediately above the
heating radius $r_h$, with convective Mach numbers $\sim 0.1$.  Convection is slightly more vigorous
in  2D, consistent with previous numerical stellar convection studies (e.g., \citealt{meakin2007}) and as
expected from the inverse turbulent cascade in 2D \citep{kraichnan1967}.

As the wind expands to larger radii, the convective motions subside, reaching
a minimum at the heating radius, and extending to regions $r< r_h$. Once
the system has reached steady-state, the magnitude of the turbulent
velocity $\Delta \mathbf{v}$ is a few $\%$ of the mean wind flow at the
sonic point, and even weaker at larger radii.   
This justifies neglecting  convection in the analytics in \S \ref{sec:analytics}.


\subsection{Time-Dependent Hydrodynamics with MESA}
\label{sec:mesa}

In this section we use the implicit hydrodynamics capabilities of the MESA stellar evolution code \citep{mesaIII} to calculate an example solution for the hydrodynamic response of a star to super-Eddington energy input.   We utilize the analytic results from \S \ref{sec:analytics} and the idealized numerical experiments in \S \ref{sec:flash} to interpret the more complete MESA models in this section.   This calculation differs from those presented in \S \ref{sec:conv} in that the latter were hydrostatic and so did not have the option of producing a wind.   The MESA models differ from the FLASH models of the previous subsection in multiple ways, perhaps most importantly by including radiation diffusion and a realistic stellar progenitor and equation of state.   The former allows us to explicitly check that neglecting radiation transport is a reasonable approximation for $\dot E_w \gtrsim L_{\rm Edd}$.        

\begin{figure*}
\centering
\includegraphics[width=8.7cm]{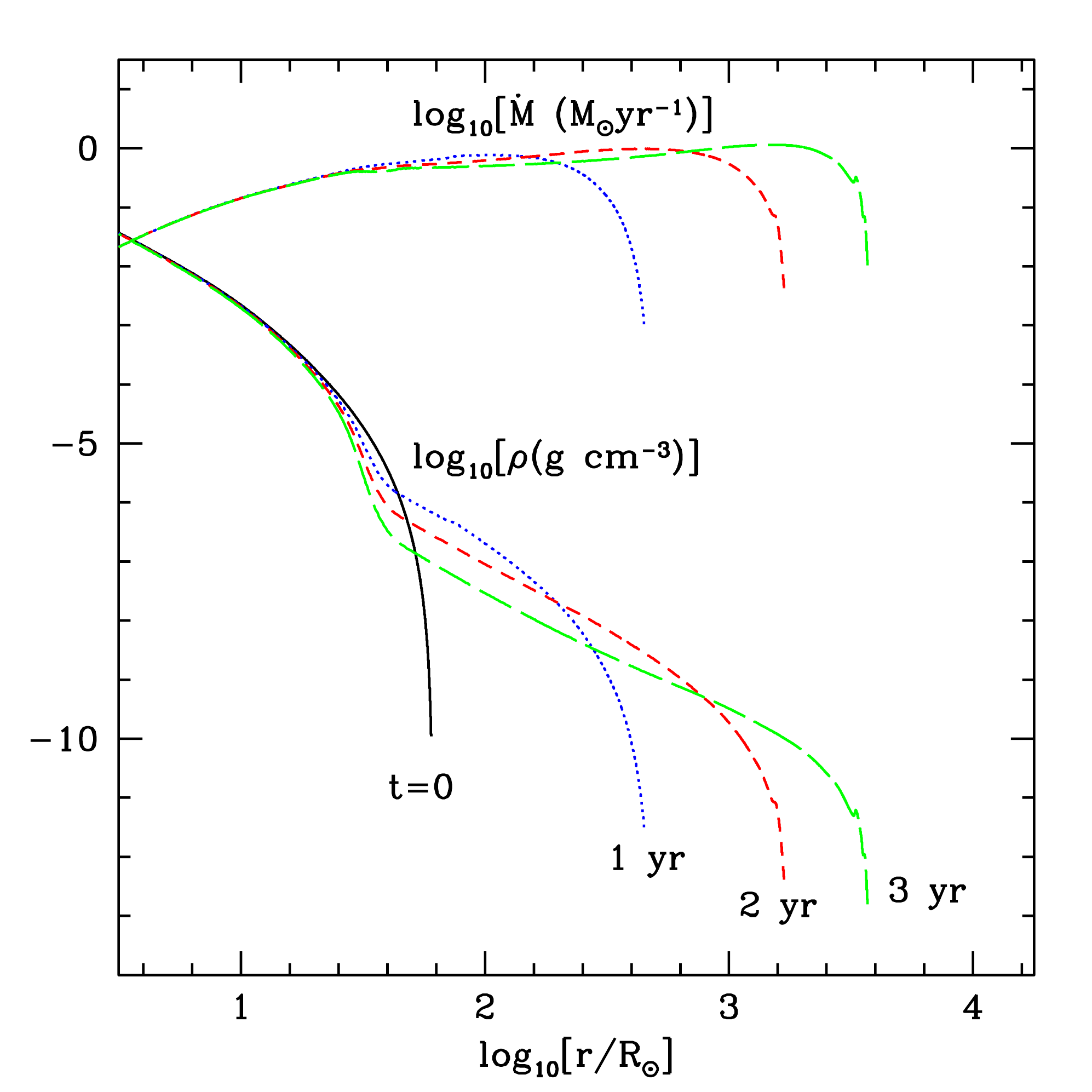}
\includegraphics[width=8.7cm]{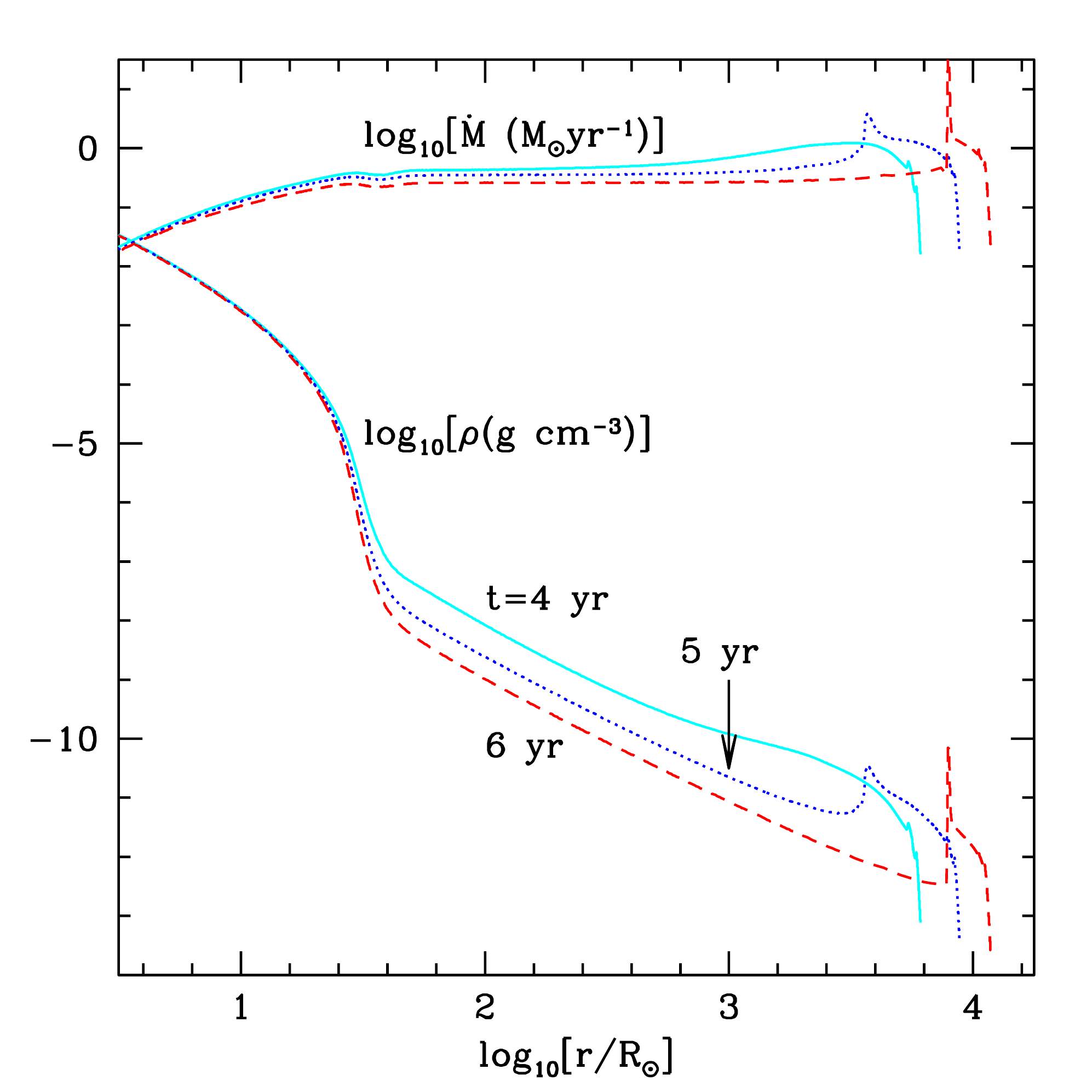}\\
\caption{Density  and mass loss rate $\dot M  = 4 \pi r^2 \rho v$ profiles in hydrodynamic MESA calculations, at different times after the onset of energy deposition with $\dot E =  10^7 L_\odot$ at $r_h \simeq 35 R_\odot$ (in a 23 $M_\odot$, $R = 60 R_\odot$, $Z = Z_\odot$ model at core He exhaustion).  The initial thermal time (eq. \ref{eq:thermal}) at $r_h$ is $\simeq 0.4$ yr, which sets the initial expansion time of the envelope.  At late times the heating has generated a quasi-steady state wind with $\dot M \simeq 0.4 \mspy$ (independent of radius over a factor of $\sim 100$ in radius).   In addition, the higher speed wind at late times shocks on the slow wind at earlier times (see the velocity profiles in Fig. \ref{fig:v-hydro}), generating a swept up shell of mass visible as the excess density/$\dot M$ at large radii.}
\label{fig:rho-hydro}
\end{figure*}

The inlists for our MESA hydrodynamic calculation are given in Appendix \ref{sec:mesa2}.  The key properties of this model include that we utilize MESA's implicit hydrodynamics capabilities, which amounts to solving the time dependent spherically symmetric hydrodynamics equations.  This includes radiation in the diffusion approximation assuming local thermal equilibrium.    We do not utilize the MLT++ convection module used in \S \ref{sec:conv} since this would over-estimate the efficacy of convective energy transport.  We also turn off MESA's wind models so that the mass on the domain is conserved.    The one significant simplification in our calculation is that we set the opacity to be electron scattering everywhere.   Calculations with a constant opacity  100 times larger yielded very similar results for the wind properties, demonstrating that radiation diffusion is not critical for the outflow properties, consistent with the arguments in \S \ref{sec:analytics}.   Our restriction to a constant opacity is primarily because of complications associated with the precipitous drop in opacity when hydrogen recombines below $\lesssim 10^4$ K.   This recombination is likely  important for understanding the outer portions of super-Eddington stellar winds, including the observational signatures of such winds.   We leave a  study of this important aspect of the problem to future work (but see \S \ref{sec:observables} for a brief discussion).     

We carried out calculations with a variety of massive stellar progenitors and energy deposition rates/locations. We present one illustrative example here.   We consider a 30 $M_\odot$, $Z = Z_\odot$ stellar model evolved to He exhaustion, at which point its mass and radius were $23 M_\odot$ and $60 R_\odot$, respectively.\footnote{These stellar parameters differ from the models in \S \ref{sec:conv} despite being for the same ZAMS stellar properties and stellar wind model.  This is in part because here we use electron scattering opacity only.  Moreover, not using MLT++ changes the stellar structure and the integrated effect of stellar winds on the model by the time of He exhaustion.}    We then deposited $\dot E = 10^7 L_\odot$  in a region centered at $\rh = 35 R_\odot$.       Given the density profile prior to heating in this stellar model, the dimensionless energy injection rate as defined for the FLASH models in \S \ref{sec:flash} is $\dot E/r_h^2 \rho(r_h) \ve(r_h)^3 \simeq 6 \times 10^{-3}$.   This is similar to the heating rate in the FLASH model L046R2.5 (1D, 2D, and 3D) shown in Figures \ref{f:flash_1d_snapshots},  \ref{f:flash_2d_snapshots}, and  \ref{f:mach_flash_snapshots}.   As time goes on, however, $\rho(r_h)$ decreases in the MESA models (as discussed below) so that the dimensionless $\dot E$ increases, making the MESA model at late times more analogous to the higher $\dot E$ FLASH models.

The left panel of Figure \ref{fig:rho-hydro} shows the density profile and mass loss rate $\dot M \equiv 4 \pi r^2 \rho v$ at early times in the response to energy deposition in the MESA models while the right panel of Figure {\ref{fig:rho-hydro} shows the same quantities during the phase when a strong steady wind is established.    Figure \ref{fig:v-hydro} shows the velocity profile at these same times.    

The total stellar mass initially exterior to the heating radius is $\sim 0.3 M_\odot$ so that the initial thermal time of the envelope is $\sim 0.4$ yr.   The stellar envelope initially expands outwards on of order the initial thermal time, driven by the excess thermal energy deposited near $\sim r_h$.   The dynamics in this phase is reminiscent of the hydrostatic models in \S \ref{sec:conv} but even at this early stage the density profile is somewhat flatter than the convective models in Figure \ref{fig:density}.   The evolution of the density profile in the MESA model is very similar to that seen in the FLASH simulations in Figure \ref{f:flash_1d_snapshots}.  The mass outflow rate in Figure \ref{fig:rho-hydro} 
is relatively independent of time at $\simeq 0.4 \mspy$, even at early times, but the velocity of the flow continues to accelerate reaching $\sim 300 \kms$ at $t = 6$ yr.    A quasi-steady outflow only develops around $t = 3$ yr.     As a result, at early times $\dot M$ is best interpreted as being due to the nearly hydrostatic expansion of the stellar envelope, rather than a wind.    During the wind phase at $t \gtrsim 4$ yr, $\dot M$  becomes nearly independent of radius over a wide range of radii, as expected for a steady wind.  

\begin{figure}
\centering
\includegraphics[width=8.5cm]{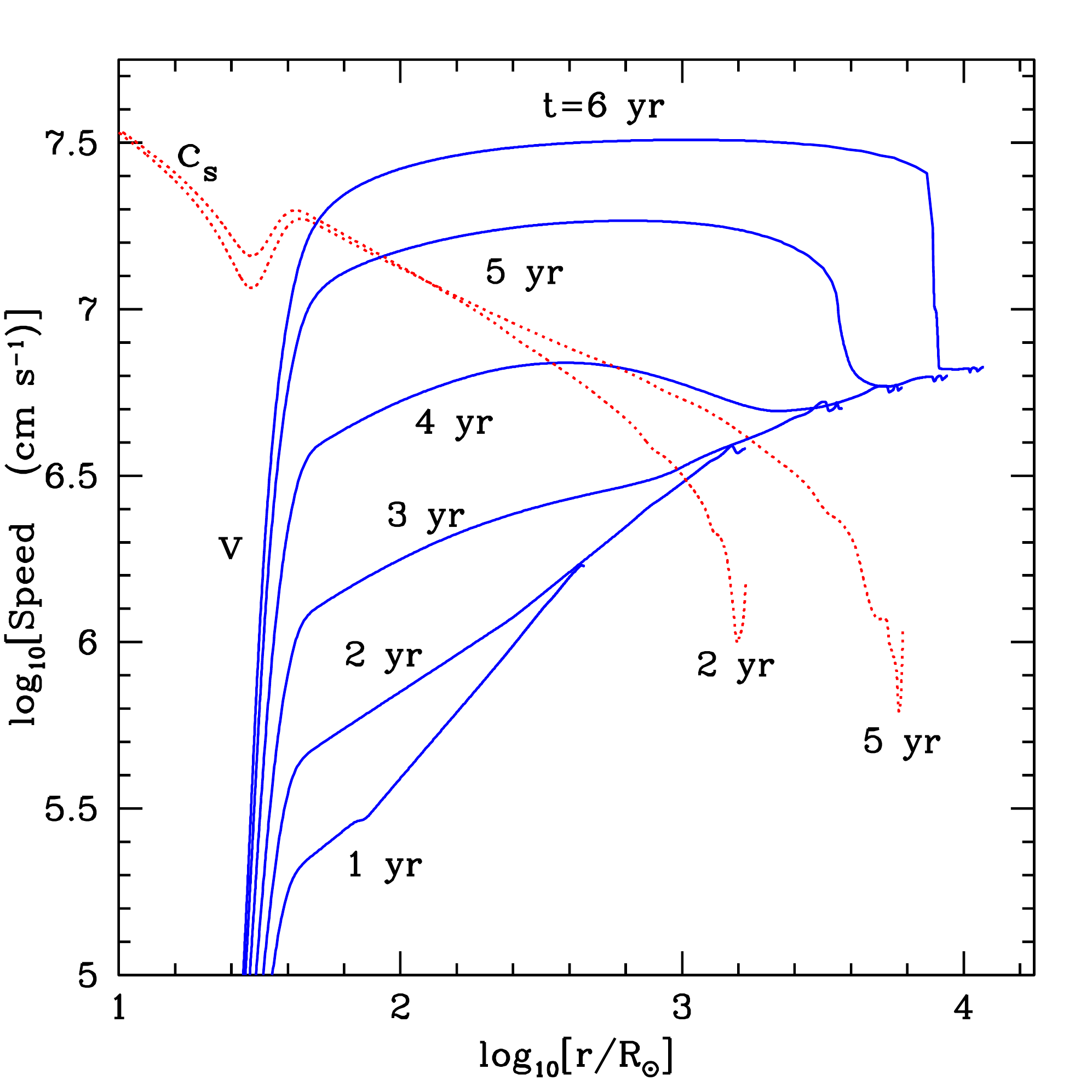}
\caption{Outflow velocity and sound speed as a function of radius in hydrodynamic MESA calculations, at different times after the onset of energy deposition with $\dot E =  10^7 L_\odot$ at $r_h \simeq 35 R_\odot$ (in a 23 $M_\odot$, $R = 60 R_\odot$, $Z = Z_\odot$ model at core He exhaustion).  As the energy deposition ejects the stellar envelope to large radii, the velocity of the outflow accelerates to $\sim 300 \kms$ and the sonic point moves in to small radii.  At late times most of the energy deposited in the stellar envelope is carried to large radii by the wind (see Fig. \ref{fig:lum-hydro}).}
\label{fig:v-hydro}
\end{figure}

\begin{figure}
\centering
\includegraphics[width=8.5cm]{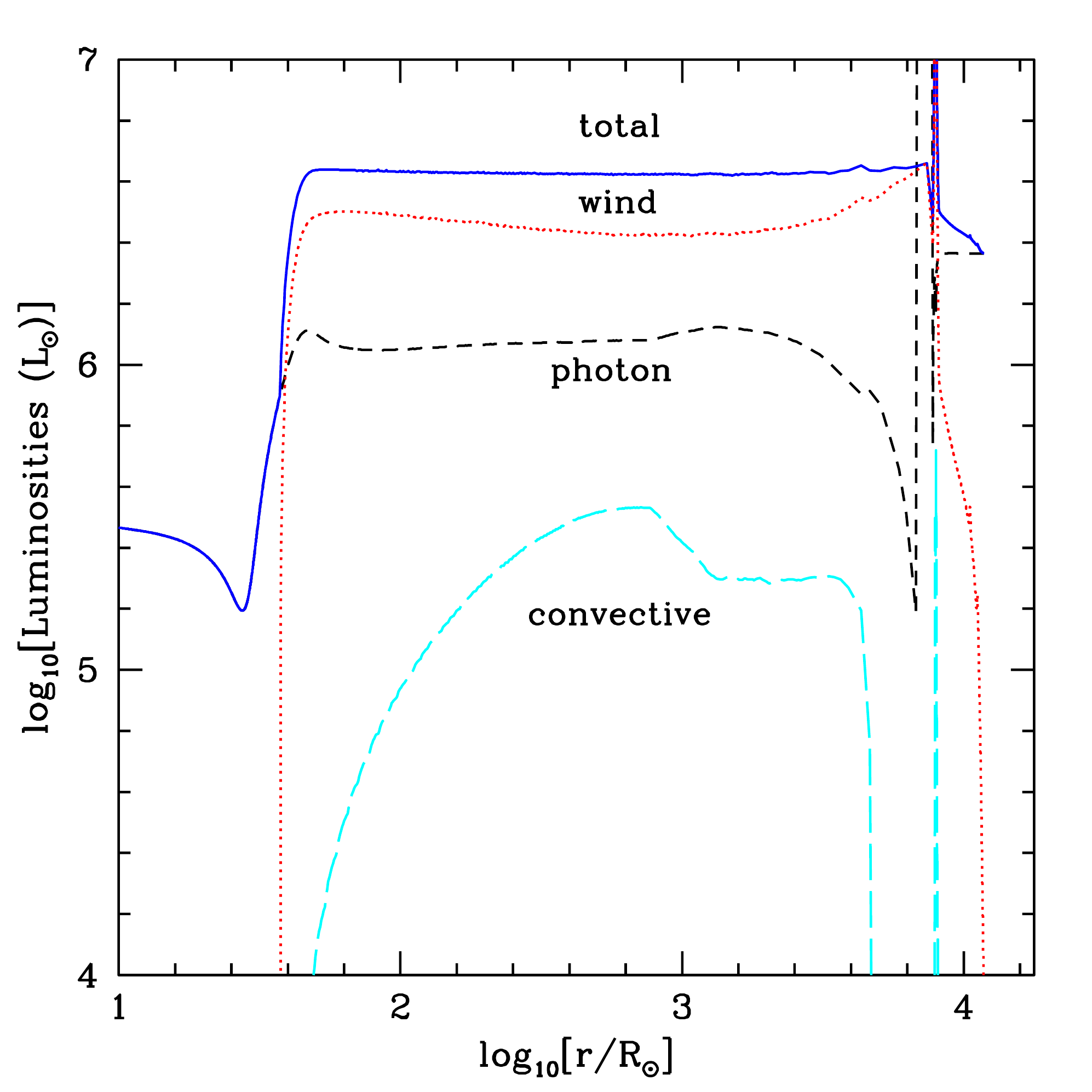}
\caption{Total, wind, photon, and convective luminosities as a function of radius in hydrodynamic MESA calculations, 6 yr after the onset of energy deposition with $\dot E =  10^7 L_\odot$ at $r_h \simeq 35 R_\odot$ (in a 23 $M_\odot$, $R = 60 R_\odot$, $Z = Z_\odot$ model at core He exhaustion).   The wind power is $\dot E_w = Be \dot M$, where $Be$ is the Bernoulli parameter.   The total power is the sum of the other three contributions.   The wind is the dominant energy transport mechanism to large radii at these late times when the outflow speed is $\sim 300 \kms$ and the sonic point of the wind has moved in to small radii (see Fig. \ref{fig:v-hydro}).  Convective energy transport is negligible, in contrast to the hydrostatic models in Figure \ref{fig:lum}.  Note also that the emergent photon luminosity at the outer edge of the wind is larger than the photon luminosity in the wind at intermediate radii.   This is due to the strong shock produced as the high speed wind at small radii encounters the slower moving shell at large radii, thermalizing the kinetic energy of the wind (Fig. \ref{fig:v-hydro} \& \S \ref{sec:shell}).}
\label{fig:lum-hydro}
\end{figure}

The increase in the wind velocity with time seen in Figure \ref{fig:v-hydro} leads to a pile up of mass at large radii as the wind runs into mass ejected at earlier times.   This accounts for the shell of matter at radii $\sim 10^{3.5-4} R_\odot$ (Fig. \ref{fig:rho-hydro}) and the strong shock evident in the velocity profile at late times (Fig. \ref{fig:v-hydro}).   At this stage of evolution, the strong stellar wind is effectively inflating a `wind bubble' inside the remnants of the outer stellar envelope.   This is not as visible in the FLASH simulations in \S \ref{sec:flash} in part because FLASH is an Eulerian code and thus matter at large radii leaves the computational domain.   

Figure \ref{fig:lum-hydro} shows the energetics of the outflow at $t = 6$ yr.   The `wind' luminosity is defined as $\dot E_w = Be \, \dot M$ while the `total' luminosity is the sum of the wind, photon, and convective contributions.\footnote{To calculate $Be$ in MESA, we calculate the gravitational potential by integrating $\nabla \phi = -GM_r/r^2$ in run\_star\_extras and add this to the enthalpy per unit mass (the sum of the internal energy per unit mass and $P/\rho$) and the kinetic energy per unit mass.}   The total outflow power is nearly independent of radius, as expected for a quasi-steady wind.   Moreover, most of the energy flux is carried by the wind, with photon diffusion contributing a total luminosity $\simeq 1.5 \, 10^{6} L_\odot$ and convection much less.   The sub-dominance of convection in Figure \ref{fig:lum-hydro} is consistent with the analytic arguments in \S \ref{sec:conv-analytics}.\footnote{In \S \ref{sec:conv-analytics} we argued that the outflow should not be convective because it is adiabatic outside the heating region.  In the MESA calculations, photon diffusion means that the outflow is not strictly adiabatic, which leads to modest energy transport by convection.}   The photon luminosity in the late steady wind stages exceeds the Eddington luminosity by a factor of $\simeq 2$, as is perfectly possible for an outflow (see \S \ref{sec:observables}).   At earlier times when the wind is much lower velocity, the photon luminosity is almost exactly the Eddington luminosity, confirming that the moderately super-Eddington photon luminosity at late times is due to the wind.     Finally,  note that the total wind power is about a factor of $\sim 2$ less than the energy input rate $\dot E = 10^7 L_\odot$.   Some of the energy is lost to work done against gravity escaping the stellar potential. 

The analytic models in \S \ref{sec:analytics} can explain many of the broad properties of the MESA simulation.   For concreteness, consider $t \sim 5$ yr when $\vc \simeq 140 \kms$, $\ve(r_h) \simeq 470 \kms$, and $M/\Me \simeq 200$.    The numerical solution at this time has $v_\infty \simeq  160 \kms$ and $\dot M \simeq 0.4 \mspy$.      The model in \S \ref{sec:analytics} predicts $\dot M \simeq 0.08 \dot M_{\rm crit} \simeq 0.5 \mspy$ and  $v_\infty \simeq \vc \simeq 140 \kms$, which are comparable to the numerical values.  
Physically, the mass loss rate corresponds roughly to $\dot E \simeq 0.5 \ve(r_h)^2 \dot M$ (Regime 2 in \S \ref{sec:analytics}).   Note that in this regime we expect the mass loss rate $\dot M$ to be relatively independent of the asymptotic velocity $v_\infty$.   This is why the mass loss rate does not change significantly in time (Fig. \ref{fig:rho-hydro}) even as the asymptotic velocity increases and the sonic point moves in from large radii (Fig. \ref{fig:v-hydro}).   Figure \ref{fig:rho-hydro} (left panel) shows that this characteristic mass-loss rate holds even when there is no super-sonic outflow.   This is because even at the early times shown here, neither photons nor convection carry a significant fraction of $\dot E$ outwards, so the majority of the energy input goes into lifting matter out to large radii.

One aspect of the MESA simulations not captured by the steady state wind models in \S \ref{sec:analytics} and \S \ref{sec:flash} is the continued evolution of the wind to higher speed and thus larger wind power even when the sonic point has moved well interior to the outer radius of the ejected mass (and so the outflow is now out of causal contact with the previously ejected matter).   There is never a true steady state.   This is not due to radiation diffusion modifying the structure of the envelope since the calculations with 100 times larger opacity yielded the same behavior.   We interpret this as due to changes in the stellar mass and envelope structure due to the outflow (the total mass ejected during the simulation is $\simeq 2.7 M_\odot$, of order 10\% of the initial mass).   The decreasing stellar mass and the ejection of the stellar envelope decrease the scale height of the envelope at the fixed heating radius $r_h$.   This in turn decreases $\Me$ and increases $\vc$, leading to a smaller sonic point radius and a more powerful outflow at late times.    It is worth stressing, however, that in reality the longer timescale evolution of the wind properties will depend in detail on how the source of heating adjusts as the envelope is ejected (there is no such back reaction in our MESA calculations).   

\section{Observational Signatures}
\label{sec:observables}

In this section we present estimates  of the observational characteristics of the wind models described earlier in this paper.   For concreteness, we assume in this section that the wind is in Regime 1 of \S \ref{sec:analytics} and use the analytic estimates of the wind properties to estimate its radiative properties.  Recall that in this regime $\dot E_w \sim \dot E$, i.e., the kinetic power in the wind is comparable to the total power $\dot E$ supplied to the stellar envelope (eq. \ref{eq:Ew-vclarge}).   The outflow in general consists of two components (\S \ref{sec:mesa}): the steady wind at small radii surrounded  by a shell  at larger radii comprised of previously ejected matter swept up by the wind.    We first estimate the observational properties of the steady wind neglecting the outer shell and then briefly discuss the impact of the outer shell.    
\vspace{-0.3cm}
\subsection{The Free Wind}
\label{sec:wind}

As the wind accelerates outwards at the sonic point $\sim r_s$, photon diffusion is initially negligible and the trapped photon energy is instead carried outwards by advection in the wind (\S \ref{sec:adiabatic}).   However, the ratio of the diffusion time $t_{\rm diff} \simeq r^2 \rho \kappa/c$ to the advection time $t_{\rm adv} \simeq r/v$ is $\propto 1/r$ for constant opacity in a steady wind.  Thus at the `diffusion radius' $r_d$, photon diffusion becomes important.   The diffusion radius can be estimated by using the fact that at $r_s$, $t_{\rm diff}/t_{\rm adv} \sim \dot E/\Ledd$, which yields
\be
r_d \sim r_s \left(\frac{\dot E}{\Ledd}\right).
\label{eq:rd}
\ee
The total bolometric luminosity radiated by the wind is set by the thermal energy content at $r_d$ and is given by
\be
\Lw \sim \Ledd \left( \frac{\dot E}{\Ledd}\right)^{1/3}.
\label{eq:lwind}
\ee
Equation \ref{eq:lwind} shows that radiated power of the wind is super-Eddington and increases with increasing heating rate $\propto \dot E^{1/3}$.   For the MESA model in \S \ref{sec:mesa}, equation \ref{eq:lwind} predicts $\Lw \sim 2.7 \Ledd$, which is similar to the factor of 2 found in the numerical solution.    Note that if density inhomogeneities in the stellar envelope and/or outflow increase the rate of photon diffusion (e.g., \citealt{shaviv2001,Owocki2004}), this effectively increases $L_{\rm Edd}$ in equation \ref{eq:lwind}, thus increasing the total power radiated by the outflow.

For $r_s \lesssim r \lesssim r_d$, the outflow is adiabatic and so the temperature is given by $T \propto \rho^{1/3} \propto r^{-2/3}$.   As a result, the temperature at $r_d$ is given by $T(r_d) \simeq T(r_s) (r_s/r_d)^{2/3}$, i.e.,
\be
T(r_d) \sim 3 \times 10^4 \, \K \ \dot E_7^{-4/15} M_{30}^{1/6} \kappa_{0.4}^{-2/3}   \left(\frac{M}{10^3 \Me}\right)^{3/20}
\label{eq:Trd}
\ee
where $\kappa_{0.4} = \kappa/0.4 \, {\rm cm^2 \, g^{-1}}$ is the opacity at $r_d$ scaled to the electron scattering opacity.   The temperature in equation \ref{eq:Trd} depends only weakly on model parameters and is sufficiently hot that the electron scattering opacity is likely a reasonable first approximation for estimating $r_d$ and $\Lw$ in equations \ref{eq:rd} and \ref{eq:lwind}.   For more quantitatively accurate estimates the diffusion radius $r_d$ can be obtained from $t_{\rm diff} \simeq t_{\rm adv}$ using a more realistic opacity $\kappa(\rho,T)$.   

The effective temperature associated with the observable emission $\Lw$ will in general be less than $T(r_d)$ in equation \ref{eq:Trd} because the optical depth at $r_d$ is still quite large $\sim c/v_\infty$.   Formally in the free steady wind, the photosphere is at a very large radius such that the time for the outflow to reach the photosphere is hundreds of years for our fiducial massive star parameters.   This implies that in practical cases of interest, the photosphere is limited by the radius the wind has reached at time $t$, $r \sim v_\infty t$.   This yields an effective temperature of
\be
T_{\rm eff} \sim  2000  \, \K \, M_{30}^{1/6} \kappa_{0.4}^{-1/6} \dot E_7^{0.017}  \left(\frac{M}{10^3 \Me}\right)^{-1/10}  t_{\rm yr}^{-1/2} 
\label{eq:teff}
\ee
where $t_{\rm yr}$ is the time since the steady outflow was initiated in years.    

The applicability of equation \ref{eq:teff} is limited by several factors.   Once $T_{\rm eff}$ falls below $\sim 5000$ K (i.e., after a few months), recombination is important and will fix the effective temperature to $\sim 5000$ K.  The photosphere will then lie inside the outer radius of the wind that is continuing to move outwards.  An additional complication is that stars with mass loss rates as large as equation \ref{eq:Mdot-vclarge} can form dust outside a radius of $\sim 10^{15}$ cm \citep{Kochanek2011}.   This implies that after of order a few-10 years dust will begin to form and will reprocess some of the wind emission to even longer wavelengths.   Prior to efficient dust formation, however, the estimates of this section demonstrate that the super-Eddington wind is likely to be a bright super-Eddington optical source.   At later times the emission will be increasingly in the infrared.   

\vspace{-0.5cm}
\subsection{The Swept-Up Shell}
\label{sec:shell}

The estimates of the previous section assume that once the wind is accelerated out through the sonic point that it continues to expand unimpeded to larger radii.    This is in fact not true in the MESA models in \S \ref{sec:mesa} (Figs \ref{fig:rho-hydro} and \ref{fig:v-hydro}).   Instead, at early times most of the energy input goes into slowly expanding the atmosphere outwards.   At later times, the density in the heating region is lower so the wind can be accelerated to higher speeds.  The end result is a comparatively slowly moving shell of matter at larger radii which confines (in 1D) the higher speed wind at late times.   The high speed wind shocks on the outer shell, converting its kinetic energy into thermal energy.   

For a shell of mass $M_{\rm shell}$ at radius $r$, the optical depth through the shell is $\sim 60 \, (M_{\rm shell}/ M_\odot) (r/10^{15} \, {\rm cm})^{-2} \kappa_{0.4}$.   At large radii dust will form (e.g., \citealt{Kochanek2011}) at which point the optical depth will be even larger than this estimate.   Thus much of the shocked wind energy may be thermalized and re-radiated (\citealt{Smith2013} argues for a model along these lines for Eta Carinae, in which much of the emission of the great eruption is powered by circumstellar interaction).   This implies that in many cases the radiated power by super-Eddington winds is likely to exceed the estimate in equation \ref{eq:lwind} and be closer to the kinetic power in the wind.   Note that this is true in the MESA model shown in Figure \ref{fig:lum-hydro}, where the emergent photon luminosity at the outer edge of the wind is larger than the photon luminosity advected out in the bulk of the wind by a factor of $\simeq 2$.   In detail, the efficiency of this ``internal shock" emission will depend on how the geometry and kinematics of the wind and the swept-up shell at large radii evolve in time.

\begin{figure*}
\centering
\includegraphics[width=8.7cm]{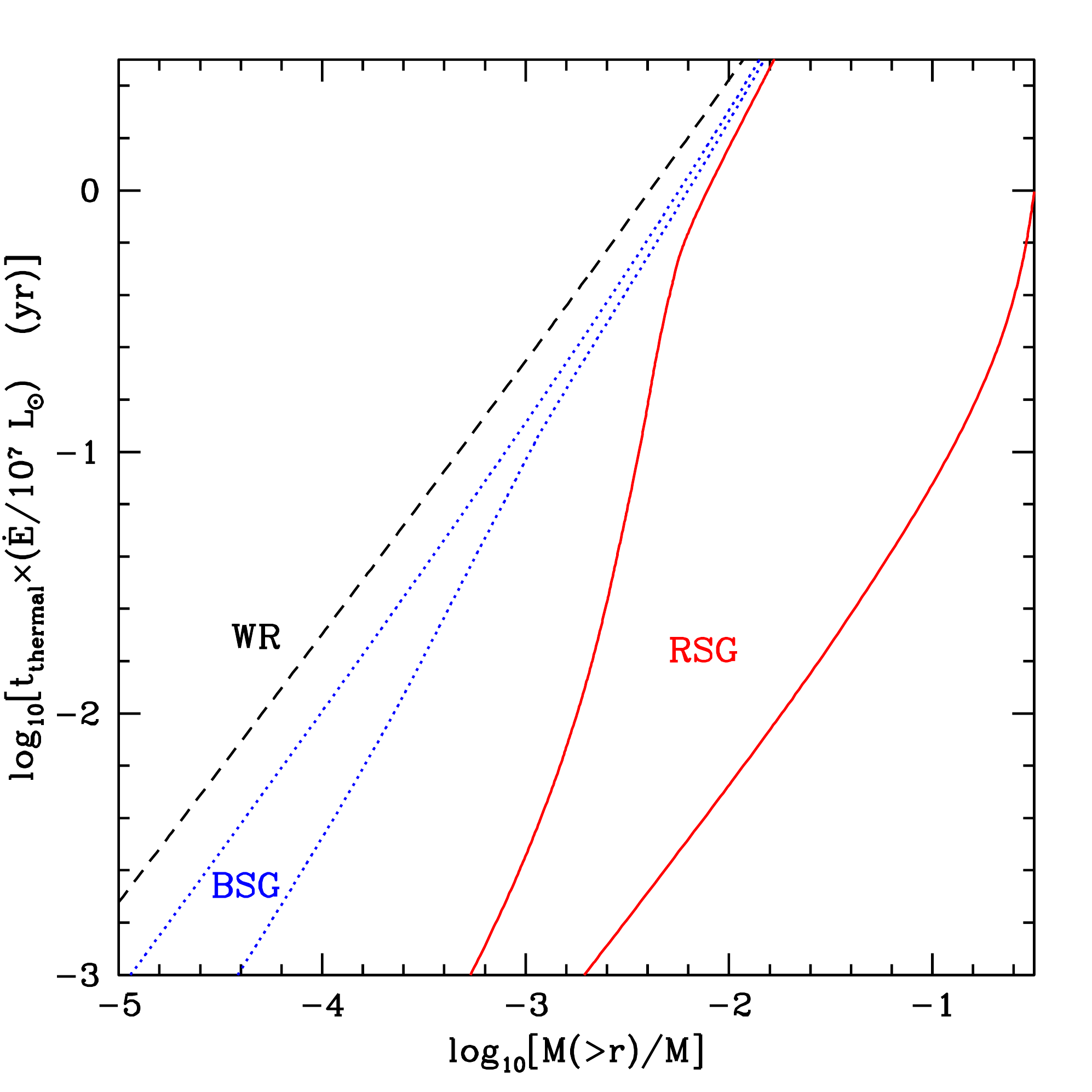}
\includegraphics[width=8.7cm]{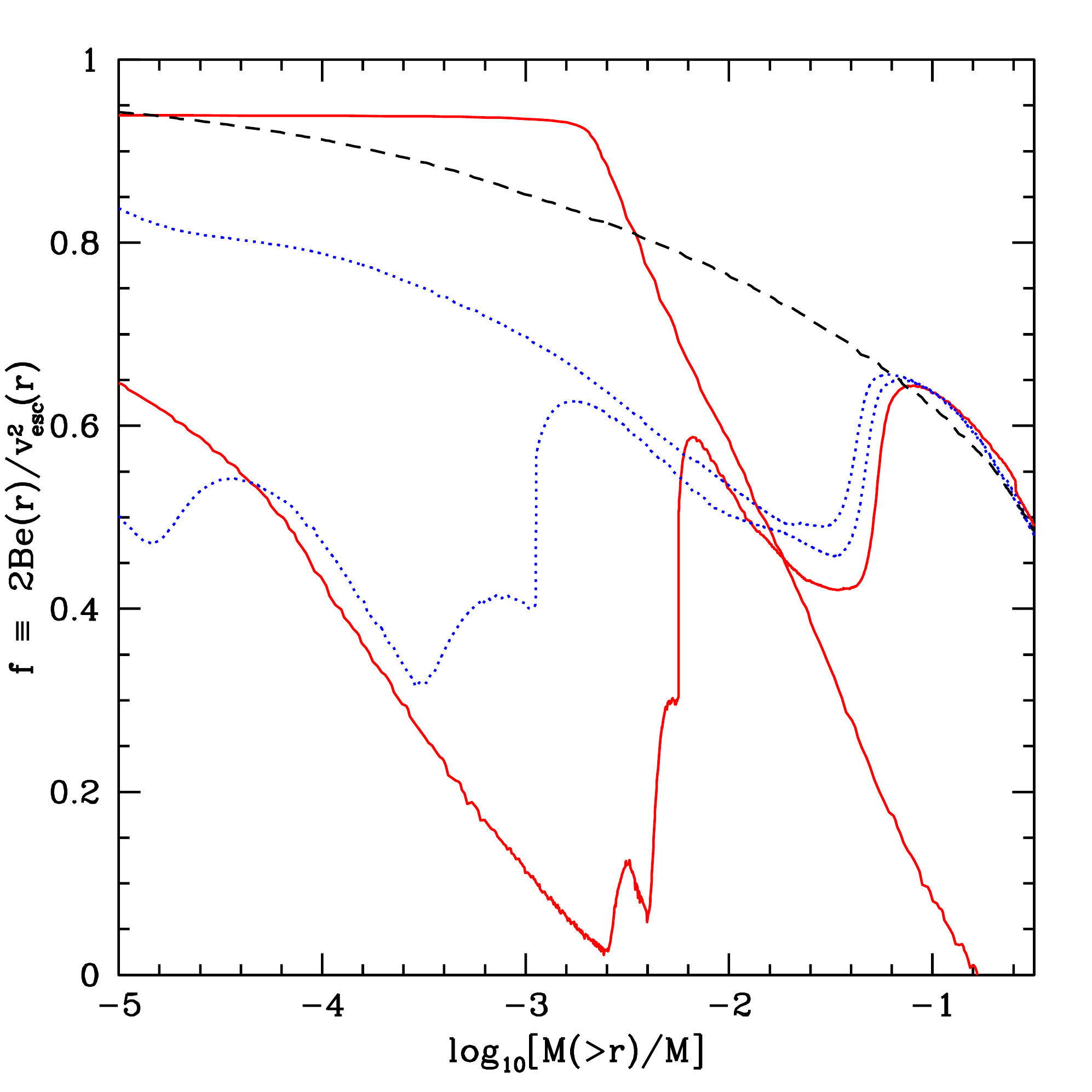}\\
\caption{{\em Left:}  Thermal time for the stellar envelope (eq. \ref{eq:thermal}) as a function of  envelope mass $M(>r)$ (mass exterior to a given radius $r$) for different massive stellar progenitors at He core exhaustion: (M/$M_\odot$,R/$R_\odot$) = (9.3,0.76), (11.1,2.6), (11.3,18.7), (11.43,275), and (19.4, 1100), from left to right in the plot, respectively.  The stellar models are labeled roughly by whether they would be classified as a Wolf-Rayet star (WR), a blue supergiant (BSG), or a red supergiant (RSG).   The thermal time sets the timescale for the envelope to respond to energy injection, and thus the time for the wind solutions calculated in this paper to apply.  It is scaled here to a heating rate of $10^7 L_\odot$, which is roughly 30 times the electron-scattering Eddington luminosity for a 10 $M_\odot$ star.     Mass ejection via super-Eddington winds on decade (e.g, Eta Carinae) or month-decade (e.g., Type IIn) timescales requires energy deposition in the outer $\sim 0.1-1\%$ of the star by mass, with the exception of the largest RSGs.   {\em Right:}   Dimensionless parameter $f \equiv 2 Be(r)/\ve^2(r)$ (eq. \ref{eq:Berh}) that characterizes the binding energy per unit mass of the stellar envelope, for the same models as in the left panel.  Lower values of $f$ in general lead to more powerful winds  (\S \ref{sec:analytics} \& Fig. \ref{fig:analytics}).    Stellar models with larger envelopes (BSG and RSG) have lower values of $f$ and are thus more prone to high speed energetic winds.}
\label{fig:thermal}
\end{figure*}

\vspace{-0.5cm}
\section{Application to Massive Stars}
\label{sec:applications}

We now briefly discuss the application of our results in the previous sections to outflows from massive stars.   To provide some context for this application, Figure \ref{fig:thermal} (left panel) shows the thermal time (defined as in eq. \ref{eq:thermal}) as a function of envelope mass for five different massive star models, from compact Wolf-Rayet-like models ($R \simeq 0.7 R_\odot$) to red super giants ($R \simeq 10^3 R_\odot$).      The models in Figure \ref{fig:thermal} are all at the end of He core fusion (see Appendix \ref{sec:mesa3}).   Recall that the thermal time scale at the heating radius sets the overall timescale on which the stellar envelope adjusts to the additional energy input and thus the timescale on which the envelope expands and a steady wind develops.   Note that for the compact progenitors in Figure \ref{fig:thermal}, $t_{\rm thermal}$ is simply  $\propto M(>r)$ because the radius $r$ does not change significantly for the envelope masses considered.   This is not true for the extended progenitors.

The right panel of Figure \ref{fig:thermal} also shows the dimensionless binding energy $f \equiv 2 Be(r)/\ve^2(r)$ (eq. \ref{eq:Berh}) of the stellar envelope.  For everything else fixed, lower values of $f$ lead to more energetic winds (\S \ref{sec:analytics} \& Fig. \ref{fig:analytics}).   Figure \ref{fig:thermal} shows that the stellar models with extended envelopes (BSG and RSG) have lower values of $f$ and are thus more prone to high speed energetic winds.    This is qualitatively consistent with the polytropic atmosphere calculation in \S \ref{sec:wind-analytics}.

\vspace{-0.3cm}
\subsection{Eta Carinae and Luminous Blue Variables}
\label{sec:lbv}

LBVs are the most dramatic manifestation of episodic mass loss in massive stars (see, e.g., \citealt{Davidson2012,Smith2014} for reviews).   The prototypical (albeit extreme) example is  Eta Carinae whose great eruption from $\sim 1840-1850$ lasted over a decade with the photon luminosity exceeding $\sim 10$ times the Eddington limit for a 50 $M_\odot$ star throughout this period.   The properties of the gas ejected during the great eruption can be  estimated from the surrounding nebula, which yields a mass, velocity, and kinetic energy of $\sim 20 M_\odot$, $\sim 500 \kms$, and $\sim 5 \, 10^{49} \erg$, respectively (\citealt{Smith2003}).   The total photon energy radiated during this period was of order, though probably somewhat less than, the kinetic energy of the ejecta.   

The physical mechanism responsible for LBV outbursts remains uncertain.   Nonetheless, the existence of super-Eddington radiative and kinetic luminosities for many dynamical times suggests that the wind models developed in this paper should be reasonably applicable.   Our models clarify the conditions required for the wind kinetic energy flux $\dot E_w$ to be of order the energy input rate $\dot E$ and larger than the radiated photon luminosity.   This condition must roughly be satisfied in LBVs given that the energy requirements to explain the $\sim 10^{50}$ ergs of kinetic energy in the outflow would go up significantly if $\dot E_w \ll \dot E$.    Our results show that $\dot E_w \sim \dot E \gtrsim L_{\rm rad}$ if the critical speed $\vc$ (eq. \ref{eq:vcrit}) is of order the escape speed at the heating location.    When this condition is satisfied, the outflow velocity is naturally a few $\vc \sim 300 \kms$ (eq. \ref{eq:v-vclarge} \& Figure \ref{fig:v-hydro}) and can be as large as that of the escape speed from the heating region.   If the current stellar mass and radius of Eta Carinae reflect the structure prior to the great eruption,  $\ve(r_h) \sim 400 \kms$ if the heating occurs near the surface, as is required to have a short thermal time (Fig. \ref{fig:thermal}).    Thus, the characteristic velocities predicted by our model are of order that observed for much of the mass in the great eruption.  We suggest that the most likely explanation for $\vc \sim \ve(r_h)$, and hence an efficient conversion of energy input to wind kinetic energy, is that the overall evolution of the density profile in response to mass loss is qualitatively similar to that found in our MESA calculations in \S \ref{sec:mesa} (Fig. \ref{fig:rho-hydro}).   In particular, if the typical mass in the stellar envelope decreases in response to mass loss, the star will naturally approach the limit in which much of the mass is ejected with $\dot E_w \sim \dot E$.

The month-decade timescale of LBV eruptions favors significant energy deposition relatively close to the stellar photosphere.    In particular, the thermal time of the stellar envelope is likely $\lesssim$ a few years to explain this evolutionary timescale.   For most stellar progenitors this requires that the excess energy input occur in the outer few percent of the stellar mass (Fig. \ref{fig:thermal}).   It is important to stress that this timescale constraint on the depth of the heating region does not limit the mass ejected to be less than a few percent of the stellar mass, because as the outer stellar envelope is shed, new matter from deeper down in the star replenishes the outer envelope.
\vspace{-0.3cm}
\subsection{Pre-Supernova Mass Loss \& Type IIn Supernovae}
\label{sec:IIn}

Up to $\sim 10\%$ of  supernova (SN) progenitors  appear to experience episodes of intense mass loss in the century to weeks leading up to core collapse (e.g., \citealt{smith2011b,Kochanek2011,kiewe2012}; see \citealt{Smith2014} for a review). We have previously proposed that powerful mass loss in the last year-decade prior to core collapse can be produced by vigorous convection in the stellar core exciting a super-Eddington wave flux that travels out into the stellar envelope and dissipates \citep{quataert2012,shiode2014}.     In addition to directly driving mass loss, wave-driven inflation of the stellar envelope could also trigger mass loss via Roche Lobe overflow in close binary systems \citep{smith2014b,mcley2014}.

The wind models developed in this paper are particularly applicable to the wave-driven mass loss mechanism since it directly corresponds to a non-local redistribution of energy in the star that can lead to super-Eddington heating rates in the stellar envelope.   Our wind models can explain the characteristic mass loss rates of $\sim 0.01-1 \mspy$ and velocities of $\sim 200-500 \kms$ inferred from many Type IIn SNe  (e.g., \citealt{kiewe2012}), if a reasonable fraction of the wind is ejected roughly in `Regime 1' of \S \ref{sec:analytics}, in which the wind energy flux $\dot E_w$ is of order the energy input rate $\dot E$ (which is the wave energy flux from the stellar core in the present context).\footnote{In some Type IIn SNe, the wind speed inferred from the narrow hydrogen lines approaches 1000 $\kms$.  In our model this requires compact progenitors with a large escape speed and wave heating at relatively low densities, i.e., close to the photosphere.   E.g., we find wind speeds of this magnitude in MESA calculations with compact BSG progenitors.}

One seemingly fine tuned aspect of this model is the need for energy to be deposited close the stellar surface in order for the thermal time to be short enough (Fig. \ref{fig:thermal}) to explain enhanced mass loss in the months-year prior to core collapse (the same concern of course arises in other models as well).   In fact, however, this is a natural property of the wave-driven mass loss model, at least for some progenitors.   The reason is that the outgoing waves in the stellar envelope are sound waves that damp primarily by radiative diffusion.   Thus they necessarily only dissipate their energy relatively close to the surface where the thermal time across the wavelength of the waves is shorter than the group travel time.  

\vspace{-0.5cm}
\section{Discussion}
\label{sec:disc}
\vspace{-0.15cm}

In this paper we have studied the physical properties of super-Eddington stellar winds.  By this we specifically mean winds in which the kinetic power approaches or exceeds the Eddington luminosity.    This work is motivated by phases in stellar evolution in which super-Eddington energy deposition can heat a region near the stellar surface, potentially driving a powerful wind.   Examples of this phase likely include classical novae, where the resulting mass loss can limit the ability of white dwarfs to approach the Chandrasekhar mass; radius expansion X-ray bursts, whose properties are important for constraining the neutron star equation of state (e.g., \citealt{Ozel2006});
luminous blue variables (LBVs), whose year-decade long outbursts may dominate the mass loss from massive stars (e.g., \citealt{Smith2006}); and enhanced pre-supernova mass loss inferred via circumstellar interaction in Type IIn supernovae.      The sources powering this excess energy deposition are varied, potentially including unstable fusion, wave redistribution of energy, large changes in opacity with temperature, or the effects of a companion in a binary system (e.g., \citealt{Schwarzschild1965,Paczynski1976,quataert2012,smith2014b}).

When the energy flux in a stellar wind is super-Eddington, the photons in the resulting outflow are  trapped with a diffusion time long compared to the outwards advection time (\S \ref{sec:adiabatic}).   In this limit the outflow is essentially adiabatic in the key region where the wind mass loss rate, velocity, etc. are set (exterior to the heating region, but interior to the sonic point).    Moreover, we have shown  analytically and numerically that convection is unimportant in quasi-steady super-Eddington stellar winds (\S \ref{sec:conv-analytics} \& \ref{sec:flash}).   This is in contrast to the roughly hydrostatic response of stars to super-Eddington heating, which inevitably drives convection that rearranges the structure of the star (e.g., \citealt{Joss1973}; see \S \ref{sec:conv}).   

The fact that photons are trapped and convection is sub-dominant simplifies the dynamics of super-Eddington stellar winds.  It implies that they can  be modeled as a $\gamma = 4/3$ radiation dominated fluid together with a physical prescription for the source of additional heating.   This is a version of \citet{Parker1958}'s solar wind model generalized to a radiation dominated fluid.     The fact that the photons are trapped also argues against radiation diffusion through a porous stellar atmosphere as being the key physics that sets the wind properties (as was suggested in previous work; e.g., \citealt{Owocki2004}): the rate of photon diffusion must be greatly enhanced for the wind to become non-adiabatic and thus for the wind properties to be significantly modified from those calculated here.    

We have analytically solved for the properties of super-Eddington winds for a simple model in which  heating at a rate $\dot E$ occurs primarily near a heating radius $r_h$.   The wind properties depend on the ratio of the escape speed at the heating radius $\ve(r_h)$ to a characteristic speed in the problem $\vc \equiv (\dot E G M/\Me)^{1/5}$ (see Fig. \ref{fig:analytics}), where $\Me/M$ is the fraction of the stellar mass near the heating radius.   For $\vc \gtrsim \ve(r_h)$, the wind kinetic power is of order the energy input rate $\dot E$.  The asymptotic wind speed is  $\sim 300 \kms$ for typical parameters relevant to massive stars (eq. \ref{eq:v-vclarge}) and  is bounded from above by $\ve(r_h)$.     In this regime, the outflow is related to what is sometimes termed the ``photon tired limit," typically taken to be when most of the photon luminosity of the star is converted into wind kinetic energy (e.g., \citealt{Owocki1997}).   In fact, the models developed here can  {exceed} this nominal limit on the wind kinetic power.   The reason is that we do not assume that the heating rate in the stellar envelope is  bounded by the photon luminosity.   This is indeed not necessarily the case for unstable thermonuclear fusion, waves excited in the stellar core heating the envelope, or heating due to an external companion (e.g., tides, common envelope).     In our models, the maximum wind power is instead bounded by the energy input rate $\dot E$ into the stellar envelope.    

For $\vc \lesssim \ve(r_h)$, the asymptotic kinetic power in the wind is less than the excess energy supplied to the stellar envelope $\dot E$.   This is because most of the energy goes into work against gravity unbinding mass from the stellar potential.   Indeed, the stellar mass loss rate in this regime is set by $\dot E \simeq 0.5 \dot M \ve(r_h)^2$ (eq. \ref{eq:Mdot-vcsmall}).   The asymptotic speed is typically $\lesssim \vc$ (eq. \ref{eq:v-vcsmall} and Fig. \ref{fig:analytics}).   

We have validated the analytic model developed here using hydrodynamic simulations with FLASH and MESA (\S \ref{sec:numerics}) (the latter utilize the new hydrodynamic capabilities of MESA; \citealt{mesaIII}).   The purpose of including simulations with  two different codes is that each explicitly tests different assumptions of the analytic model.   In particular, the FLASH simulations use a simple $\gamma = 4/3$ equation of state and do not include radiation diffusion but are multi-dimensional while the MESA simulations are one-dimensional but include radiation diffusion and a realistic equation of state and stellar progenitor.   

The key result of the FLASH simulations is that the steady state solutions of the model analytic problem in multiple dimensions  (largely 2D, but one 3D simulation) agree remarkably well with the analytic results, and with one-dimensional simulations (Fig. \ref{f:flash_wind_steady}).   This is fundamentally because convection can be important in the initial hydrostatic expansion of a stellar envelope in response to heating, but it is unimportant in the resulting steady state wind that develops (see \S \ref{sec:conv-analytics} \& Fig. \ref{f:flash_2d_snapshots}).   

Our MESA hydrodynamic simulations model super-Eddington energy input into the envelope of a massive star.  During the first  thermal time, the envelope mass at the heating location is so large that  there is not  a strong supersonic wind:  the envelope is essentially hydrostatically lifted to large radii (Figs. \ref{fig:rho-hydro} \&  \ref{fig:v-hydro}).   As the mass in the stellar envelope decreases, however, a supersonic wind quickly forms and accelerates outwards, effectively inflating a stellar wind bubble inside the slower matter ejected at earlier times.   
The wind in this phase has a mass loss rate and velocity similar to that predicted by the analytic models in \S \ref{sec:analytics}.   This explicitly demonstrates that radiation diffusion does not strongly affect the wind properties when the kinetic power in the wind is of order or larger than the Eddington luminosity.   

Although the calculations we have presented are general and potentially applicable to a range of astrophysical environments, we have in particular highlighted their application to powerful outflows from massive stars (\S \ref{sec:applications}).   This includes both LBVs outbursts such as Eta Carinae's great eruption and the large mass loss rates that precede some core-collapse supernovae.

LBV outbursts are characterized by a super-Eddington photon luminosity and an outflow with a super-Eddington kinetic power (e.g., \citealt{Smith2003}).  The fact that this activity is on for many dynamical times argues against a pure shock-mediated phenomena and in favor of an energy source that drives a continuous wind.   Although the ultimate energy source for LBV eruptions is poorly understood, we argue that our models can explain the characteristic mass-loss rates of $\sim 1 \mspy$ and velocities of several hundred $\kms$ of much of the ejected mass (\S \ref{sec:lbv}).  We speculate that the evolution during an eruption shares some qualitative similarities with the MESA models in Figures \ref{fig:rho-hydro} \& \ref{fig:v-hydro}:  at early times much of the input energy goes into inflating the stellar envelope.  Only then can the outflow be accelerated to high speeds.    

Our super-Eddington wind models  produce a super-Eddington photon luminosity like that observed in LBV eruptions and some pre-SN outbursts from massive stars (e.g., \citealt{smith2011a,ofek2013}).   The luminosity of the  steady wind freely expanding into a vacuum  is given by $\sim L_{\rm Edd} (\dot E/L_{\rm Edd})^{1/3}$ (eq. \ref{eq:lwind}).   However, in our MESA calculations, the strong shock driven as the higher speed wind at late times encounters the slower, previously ejected envelope can thermalize much of the wind kinetic power.  Depending on the exact wind kinematics as a function of time, this may substantially increase the photon luminosity of the wind (\S \ref{sec:shell} \& Fig. \ref{fig:lum-hydro}).    More realistic radiation transfer calculations of this process would be particularly valuable in quantitatively connecting the  models developed here to observations.

Throughout this work we have emphasized the essentially hydrodynamic character of super-Eddington stellar winds, in which radiation transfer is not dynamically that important because the photons are trapped and advected with the fluid.   We  acknowledge that although this is true in the models we have developed, it is possible that more realistic calculations will show that the outflow is sufficiently inhomogeneous to substantially increase the rate at which photons diffuse through the outflow (e.g., \citealt{shaviv2001, Owocki2004}).   Understanding this more quantitatively will ultimately require multi-dimensional radiation (magneto)-hydrodynamical simulations (e.g., \citealt{Jiang2015}).  

One particularly interesting extension of the work described in this paper is to the possibility of outflows generated when large changes in opacity with temperature near the surface of a star cause the luminosity to suddenly be super-Eddington.    If this occurs sufficiently close to the photosphere, convection may be unable to carry  outwards the energy  generated in the stellar interior.  In hydrostatic stellar models, this leads to gas pressure and density inversions (e.g., \citealt{Joss1973,mesaII}).   In hydrodynamic models, however, a wind may develop (e.g., \citealt{Kato1994,Eichler1995}).  
For this application, our calculations need to be extended to take into account the temperature dependence of the opacity and  the fact that the flux may be super-Eddington over only a modest  range in temperature (e.g., at the iron opacity bump). 

In future work it would also be  valuable to extend our analysis to lower mass outflow rates.  In particular, when the kinetic power of the wind is below the Eddington luminosity, it is no longer correct to assume -- as we have done --  that the wind is adiabatic between the heating region and the sonic point (\S \ref{sec:adiabatic}).   Understanding outflows with lower mass loss rates  thus requires a more careful treatment of the effects of photon diffusion.

\vspace{-0.5cm}
\section*{Acknowledgments}
\vspace{-0.1cm}

We thank Lars Bildsten, Josiah Schwab, and Nathan Smith for  useful conversations, and Nathan Smith for valuable comments on an initial draft of the paper.   This work was supported in part by NSF grant AST-1205732.  EQ was also supported by a Simons Investigator award
from the Simons Foundation and the David and Lucile Packard Foundation. 
RF acknowledges support from the University of California Office of the President, and
from NSF grant AST-1206097.
DK is supported in part by a Department of Energy Office of Nuclear Physics Early
Career Award, and by the Director, Office of Energy Research, Office of High
Energy and Nuclear Physics, Divisions of Nuclear Physics, of the U.S.
Department of Energy under Contract No. DE-AC02-05CH11231.
The software used in this work was in part developed by the DOE NNSA-ASC OASCR Flash Center at the
University of Chicago.
This research used resources of the National Energy Research Scientific Computing
Center (NERSC), which is supported by the Office of Science of the U.S. Department of Energy
under Contract No. DE-AC02-05CH11231. FLASH computations were performed at 
\emph{Carver} and \emph{Hopper} (repo2058).

  \vspace{-0.5cm}
\bibliography{winds2,rodrigo}
\appendix
\section{Stellar Models}

Our stellar models are constructed using the MESA stellar evolution code, version 7664.   This version includes several additional boundary conditions for the full hydrodynamic evolution that were useful for the wind solution in \S \ref{sec:mesa} (these boundary conditions are not in MESA release 7624).    The models use the following inlist controls file.   All runs use ${\rm inlist\_massive\_defaults}$ in addition to the specific flags given below.   
\vspace{-0.5cm}
\subsection{Hydrostatic Models With Heating From \S \ref{sec:conv}}
\label{sec:mesa1}
We first run a model to He core exhaustion using
{\small \begin{lstlisting}
&star_job

      create_pre_main_sequence_model = .true.

&controls

      okay_to_reduce_gradT_excess = .true. 

      initial_mass = 30 
      Zbase = 0.02

      varcontrol_target = 1d-3
         
      Dutch_scaling_factor = 1.9
      Dutch_wind_lowT_scheme = `de Jager'
      Hot_wind_scheme = `Dutch'
 
      xa_central_lower_limit_species(1) = `he4'
      xa_central_lower_limit(1) = 1d-5
\end{lstlisting}}
\noindent The wind parameter ${\rm Dutch\_scaling\_factor}$ was chosen as above to end up with a compact stellar model at He core exhaustion, since this most effectively demonstrates the effects of extra energy deposition.   We then used the output model above as an input model and evolved it using the extra\_energy subroutine in ${\rm run\_star\_extras.f}$ (described below in \S \ref{sec:mesa-htg}) with the same ${\rm \&controls}$ as above except
{\small \begin{lstlisting}
&star_job

       set_initial_age = .true.
       initial_age = 0

       set_initial_dt = .true.
       years_for_initial_dt = 0.001
       
       change_v_flag = .true.
       change_initial_v_flag = .true.
       new_v_flag = .true.           

&controls

       use_other_energy = .true.
       x_ctrl(1)=3d6
       x_ctrl(2)=2.   
       x_ctrl(3)=0.3
       x_ctrl(4)=0.1  
       x_ctrl(5)=10.  
	
       varcontrol_target = 1d-4
        
       Dutch_scaling_factor = 0.

       max_age  = 100.

\end{lstlisting}}
\noindent The choice of max\_age is arbitrary.   The models were stopped roughly after the times shown in Figures \ref{fig:density} \& \ref{fig:lum}.

\vspace{-0.3cm}
\subsection{Hydrodynamic Models with Heating From \S \ref{sec:mesa}}
\label{sec:mesa2}
We first run a model to H core exhaustion using the same inlist parameters as in the first model of Appendix \ref{sec:mesa1} except with
{\small \begin{lstlisting}
&controls
      okay_to_reduce_gradT_excess = .false.
      
      mixing_length_alpha = 1.89
\end{lstlisting}}
\noindent The choice of a different mixing\_length\_alpha is unimportant.   We then run a model to He core exhaustion using
{\small \begin{lstlisting}
&star_job
     
      change_v_flag = .true.
      change_initial_v_flag = .true.
      new_v_flag = .true.	    
           
&controls

      okay_to_reduce_gradT_excess = .false. 

      initial_mass = 30  
      Zbase = 0.02

      varcontrol_target = 1d-3
     
      mixing_length_alpha = 1.89

      Dutch_scaling_factor = 1.9
      Dutch_wind_lowT_scheme = `de Jager'
      Hot_wind_scheme = `Dutch'
 
      xa_central_lower_limit_species(1) = `he4'
      xa_central_lower_limit(1) = 1d-5
      
      use_simple_es_for_kap = .true.     
\end{lstlisting}}
\noindent The key differences relative to the model in Appendix \ref{sec:mesa1} are the use of electron scattering opacity only and the absence of MLT++.  Use of MLT++ in the hydrodynamics calculation would artificially suppress the generation of a wind.   For consistency, we thus evolved the progenitor model prior to energy deposition without MLT++.   We then used the model above as an input model and excised the core using
{\small \begin{lstlisting}
&star_job
	
      remove_initial_center_by_mass_Msun = 13

&controls

      max_model_number = 1
\end{lstlisting}}
\noindent The removal of the core allows the subsequent hydrodynamic calculation to focus on the envelope where the energy deposition occurs.    Finally, we used the output model above with an excised core as an input model and evolved it using MESA's implicit hydrodynamics solver.   We used the extra\_energy subroutine in ${\rm run\_star\_extras.f}$  (described below in \S \ref{sec:mesa-htg}) with the same ${\rm \&controls}$ as above except
{\small \begin{lstlisting}
&star_job

     relax_to_this_tau_factor = 1d-4
     dlogtau_factor = .1
     relax_initial_tau_factor = .true.

     change_E_flag = .false.  
     new_E_flag = .false.

     set_initial_age = .true.
     initial_age = 0

     set_initial_dt = .true.
     years_for_initial_dt = 0.001

&controls

     okay_to_reduce_gradT_excess = .false.  

     use_Type2_opacities = .false.
     use_simple_es_for_kap = .true.
        
     x_ctrl(1)=1d7 
     x_ctrl(2)=35.  
     x_ctrl(3)=4.   
     x_ctrl(4)=0.1   
     x_ctrl(5)=10.   

     mixing_length_alpha = 1.89
     min_T_for_acceleration_limited_conv_velocity = 0
 
     varcontrol_target = 1e-4
     
     Dutch_scaling_factor = 0.0  

     use_ODE_var_eqn_pairing = .true.
     use_dvdt_form_of_momentum_eqn = .true.
     use_dPrad_dm_form_of_T_gradient_eqn = .true.

     use_compression_outer_BC = .true.
     which_atm_option = `simple_photosphere'
     use_zero_dLdm_outer_BC = .true.

     use_artificial_viscosity = .true.
     shock_spread_linear = 0.
     shock_spread_quadratic = 0.01 

     mesh_delta_coeff = 0.8
     min_dq = 1d-10
     max_center_cell_dq = 5d-6
     max_surface_cell_dq = 1d-10
     log_tau_function_weight = 100
     log_kap_function_weight = 100
      
     newton_iterations_limit = 7
     iter_for_resid_tol2 = 4
     tol_residual_norm1 = 1d-8
     tol_max_residual1 = 1d-7

     tiny_corr_coeff_limit = 999999
     newton_itermin_until_reduce_min_corr_coeff = 999999
         
     max_age = 350.
\end{lstlisting}}
\noindent The choice of max\_age is arbitrary.   The models were stopped roughly after the times shown in Figures \ref{fig:rho-hydro}-\ref{fig:lum-hydro}.     

For the model with higher opacity noted in \S \ref{sec:mesa}, we used  opacity\_factor = 100 in addition to the above flags.  

\vspace{-0.4cm}
\subsection{Models From \S \ref{sec:applications}}
\label{sec:mesa3}
These models were run to He core exhaustion using the identical inlists as in the first part of Appendix \ref{sec:mesa1} (the models without heating) and ${\rm Dutch\_scaling\_factor}$ = 2.5, 1.9, 1.75, 1.7, and 0.8 for the models in Figure \ref{fig:thermal} (left to right).   Note that decreasing  or increasing varcontrol\_target  or changing the overshoot/mixing parameters can slightly change the time to He core exhaustion and thus the stellar mass at that time (the latter because of the effects of stellar winds).   This can change whether the model is a BSG, RSG, or WR star in Figure \ref{fig:thermal}.  However, models with different parameters but similar radii at He core exhaustion have similar thermal profiles so the thermal profiles in Figure \ref{fig:thermal} are more robust than the specific model parameters that produce them.
\vspace{-0.4cm}
\subsection{Extra Energy Deposition}
\label{sec:mesa-htg}
Our model for extra energy deposition uses the extra\_controls hook to call a subroutine that deposits a time-independent source of heating via other\_energy as follows:  we deposit an energy per unit time x\_ctrl(1)*$L_\odot$ as a Gaussian centered at radius x\_ctrl(2)*$R_\odot$ with dispersion x\_ctrl(3)*$R_\odot$. The Gaussian is cutoff below radius x\_ctrl(4)*$R_\odot$ (to avoid problems if there is energy deposition  too deep in the core in some calculations).   Finally, we multiply the heating by tanh(star\_age*x\_ctrl(5)) to ensure that a sudden onset of super-Eddington heating does not cause numerical problems.

\end{document}